\documentclass{scrartcl}
\usepackage{amsmath}
\usepackage{amssymb}

\usepackage{bm}

\usepackage[utf8]{inputenc}
\usepackage[T1]{fontenc}
\usepackage{mathptmx}
\usepackage{etoolbox}
\usepackage{subcaption}
\usepackage{nicefrac}

\makeatletter
\def\@email#1#2{%
 \endgroup
 \patchcmd{\titleblock@produce}
  {\frontmatter@RRAPformat}
  {\frontmatter@RRAPformat{\produce@RRAP{*#1\href{mailto:#2}{#2}}}\frontmatter@RRAPformat}
  {}{}
}%
\makeatother

\usepackage{natbib}
\usepackage{graphicx}
\usepackage{color}
\usepackage{overpic}
\usepackage{lmodern}

\newcommand{\bvec}[1]{\boldsymbol {#1}}
\newcommand{\rmd}{\text{d}}

\newcommand{\rmi}{\text{i}}

\newcommand{\dz}{\;\text{d} z}

\newcommand{\dx}{\;\text{d} x}

\newcommand{\Ro}{\text{Ro}}
\renewcommand{\Re}{\text{Re}}
\newcommand{\Rom}{\text{Ro}^{-1}}
\newcommand{\Rem}{\text{Re}^{-1}}

\newcommand{\ut}{\tilde{u}}
\newcommand{\vt}{\tilde{v}}

\usepackage{stmaryrd}
\newcommand{\jump}[1]{\llbracket #1 \rrbracket }

\renewcommand{\ell}{L}

\begin{document}


\title{Attached and separated rotating flow over a finite height ridge}
\author{S. Frei\thanks{Department of Mathematics \& Statistics, University of Konstanz, Universit\"atsstr. 10, 78457 Konstanz, Germany, stefan.frei@uni-konstanz.de}\,, 
E. Burman, E. Johnson\thanks{Department of Mathematics, University College London, UK, 
e.burman@ucl.ac.uk, e.johnson@ucl.ac.uk}}
\date{\today}
             
             \maketitle

\begin{abstract}
This paper discusses the effect of rotation on the boundary layer in high Reynolds number flow over a ridge using a numerical method based on stabilised finite elements that captures steady solutions up to Reynolds number of order $10^6$.   The results are validated against boundary layer computations in shallow flows and for deep flows against  experimental observations reported in Machicoane {\em et al.} (Phys. Rev. Fluids, 2018). In all cases considered the boundary layer remains attached, even at large Reynolds numbers, provided the Rossby number of the flow is sufficiently small. At any fixed Rossby number the flow detaches at sufficiently high Reynolds number to form a steady recirculating region in the lee of the ridge. At even higher Reynolds numbers no steady flow is found.  This disappearance of steady solutions closely reproduces the transition to unsteadiness seen in the laboratory.    
\end{abstract}

\section{Introduction}

\citeauthor{MasonS79} \cite{MasonS79} observe that topography with horizontal length scales of the order of a few kilometres is often sufficiently steep to induce flow separation in the atmosphere and that such flows might be expected to produce significant changes in the momentum exchange between the atmosphere and the surface. In addition to exerting a form drag on the fluid, separated flow over topography leads to enhanced mixing in the lee of the topography, affecting the lee-wave field and may make a significant contribution to the overall vertical mixing in the region \cite{RichardsSHCH92}. More recently, \citeauthor{Machicoane18}~\cite{Machicoane18} present theoretical and experimental results for flow past a horizontal cylinder in horizontal motion relative to a frame rotating rapidly about a vertical axis. Their work discusses in detail the inertial lee-wave field generated by the cylinder, extending significantly the results of \cite{Johnson82a}, and also includes a discussion of experimental results for the onset, with increasing translation speed of the cylinder, of separation and subsequent  unsteadiness in the flow.  They note that rotation strongly suppresses separation and suggest, from the experimental results, that with increasing flow speed the flow changes directly from being steady and attached to being unsteady and detached as the product of the Rossby and Reynolds numbers ($Ro = U/2\Omega L$, $Re =\nu/UL$ for flow speed $U$, rotation rate $\Omega$, kinematic viscosity $\nu$ and typical horizontal scale $L$) for the flow passes through $275\pm25$ . This contrasts with non-rotating flow where separation occurs at $Re\approx3$ but wake instability at $Re\approx47$, \cite{ Williamson1996, Machicoane18}. The purpose of the present work is to examine this lee separation in rotating flows in greater detail.

To obtain the critical Reynolds numbers for this nonlinear flow over a wide range of Rossby numbers, a stable numerical method is required, which is robust independently of the Reynolds number.  For this purpose, we employ a stabilised equal-order finite element method using \textit{Continuous Interior Penalty Stabilisation}, going back to Douglas \& Dupont~\cite{DouglasDupont1976}. Concerning fluid equations, it has first been used by Burman, Hansbo \& Fern\'andez for the Stokes, Darcy and Oseen equations~\cite{BurmanHansbo2006, Burmanetal2006} and later for more complex problems, such as flows at high Reynolds numbers~\cite{Tongetal2022}, including turbulent flows~\cite{Mouraetal2022}, as well as fluid-structure interactions~\cite{Fernandez2013, BurmanFernandezFrei2020}.

The structure of this article is as follows. Section~\ref{sec:eq} introduces the geometry and the governing equations. Section~\ref{S:bl} introduces a streamfunction formulation for finite height ridges and obtains the outer and inner, boundary-layer solution at large Reynolds number for shallow flow.The numerical method for the boundary layer solution is given in Appendix \ref{S:blnumerics}.  Section~\ref{sec:num:shallow} introduces the numerical method, showing that for attached flow the results for increasing Reynolds number converge towards the boundary layer solutions, but also capturing steady separated flows at larger Rossby numbers. Section~\ref{sec:num:deep} extends the results to deep flow allowing inertial waves to appear in the outer flow but showing the same qualitative behaviour of separation and unsteadiness as shallow flows, albeit at smaller Reynolds numbers. Section~\ref{sec:num:cyl} investigates deep rotating flow over a horizontal cylinder, the geometry of \citeauthor{Machicoane18}~\cite{Machicoane18}, and shows that the numerical simulations capture much of the experimental observations. Section~\ref{sec:concl} gives brief conclusions.

\section{Governing equations}
\label{sec:eq}

\begin{figure}[t]
\centering
\begin{picture}(0,0)%
\includegraphics{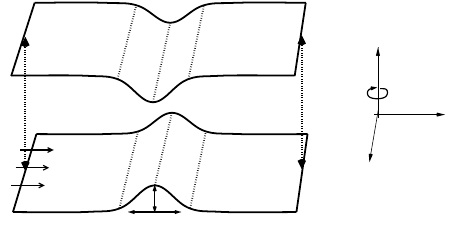}%
\end{picture}%
\setlength{\unitlength}{1243sp}%
\begingroup\makeatletter\ifx\SetFigFont\undefined%
\gdef\SetFigFont#1#2{%
  \fontsize{#1}{#2pt}%
  \selectfont}%
\fi\endgroup%
\begin{picture}(11552,5694)(1471,-6691)
\put(9316,-3481){\makebox(0,0)[lb]{\smash{{\SetFigFont{8}{9.6}{\color[rgb]{0,0,0}2*D}%
}}}}
\put(5536,-6091){\makebox(0,0)[lb]{\smash{{\SetFigFont{8}{9.6}{\color[rgb]{0,0,0}h}%
}}}}
\put(12916,-3976){\makebox(0,0)[lb]{\smash{{\SetFigFont{8}{9.6}{\color[rgb]{0,0,0}x}%
}}}}
\put(10981,-2041){\makebox(0,0)[lb]{\smash{{\SetFigFont{8}{9.6}{\color[rgb]{0,0,0}z}%
}}}}
\put(1486,-5236){\makebox(0,0)[lb]{\smash{{\SetFigFont{8}{9.6}{\color[rgb]{0,0,0}U}%
}}}}
\put(2296,-3706){\makebox(0,0)[lb]{\smash{{\SetFigFont{8}{9.6}{\color[rgb]{0,0,0}2*D}%
}}}}
\put(10666,-5371){\makebox(0,0)[lb]{\smash{{\SetFigFont{8}{9.6}{\color[rgb]{0,0,0}y}%
}}}}
\put(5131,-6676){\makebox(0,0)[lb]{\smash{{\SetFigFont{8}{9.6}{\color[rgb]{0,0,0}L}%
}}}}
\end{picture}%

\caption{Flow over a ridge. We assume a constant inflow $U$ in horizontal direction, while the geometry rotates around the vertical $z$-axis.\label{fig_bump} }
\end{figure}
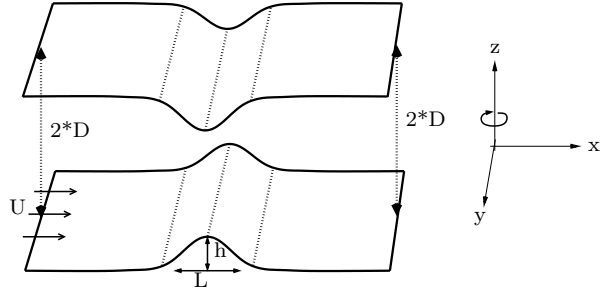

Consider a fluid of constant density $\rho$ and viscosity $\nu$ in solid body rotation about a vertical axis at angular velocity $\Omega$. Introduce Cartesian axes $Ox^*y^*z^*$ fixed in the fluid with $Oz^*$ vertical. Let the fluid lie above a horizontal boundary perturbed by a ridge of height $h$ and horizontal scale $L$, given by $z^*=h\; b(x^*/L)$, where the function $b$ gives the ridge profile and has a maximum value unity. Let the fluid be bounded above at height $z^*=2D$ by a horizontal boundary perturbed downwards by a complementary ridge so that the flow geometry is symmetric about the line $z^*=D$ and the motion has aspect ratio $\delta=D/L$ (Figure~\ref{fig_bump}).

Now suppose the fluid is set in steady motion, relative to the rotating frame, so that at large distance, outside viscous layers on the boundaries, the flow is uniform with constant speed $U$ in the positive $x^*$ direction. The Coriolis force associated with this uniform flow is balanced by a constant pressure gradient of magnitude  $2 \rho\Omega U $ in the negative $y^*$ direction. In the absence of the ridge ($h=0$), and far upstream of the ridge when $h>0$, the flow is the usual Ekman layer flow with a boundary layers of thickness $(\nu/\Omega)^{1/2}$ on the upper and lower boundaries. The flow is independent of the coordinate $y^*$ and so non-dimensionalising lengths on $L$, velocities on $U$ and the pressure on $\rho U^2$ gives the two-dimensional governing equations
\begin{subequations}
\label{maineq}
\begin{align}
 uu_x+wu_z - \Rom v &= -p_x +\Rem\nabla^2 u, \label{momx1}\\
 uv_x+wv_z + \Rom (u-1) &= \Rem\nabla^2 v,\label{momy1}\\
 uw_x+ww_z  &= -p_z +\Rem\nabla^2 w,\label{momz1}\\
 u_x+w_z&=0, \label{cty}
\end{align}
\end{subequations}
with the constant in \eqref{momy1} giving the constant far-field pressure gradient in the $y$-direction. On the top and bottom boundaries the flow satisfies the viscous boundary conditions that
the velocity vanishes., i.e. $u=v=w=0$.
The two non-dimensional parameters appearing are the Reynolds number $\Re =UL/\nu$, measuring the ratio of inertial to viscous forces, and the Rossby number $\Ro=U/2\Omega L$, measuring the ratio of inertial to Coriolis forces. Since the nonlinear inertial terms are always significant in the flows discussed below, $\Re$ and $\Ro$ are taken, together with the aspect ratio $\delta$, as the fundamental parameters. The Ekman number $E=\nu/2\Omega D^2= \Ro/\delta^2\Re$ appears naturally with $E^{1/2}$ measuring the ratio of the Ekman layer thickness to the fluid depth and is treated as derived from the fundamental parameters. 

Integrating \eqref{cty} across the fluid depth gives the dimensional result
\begin{equation}
 \int u^* \dz^* = 2UD, \label{mass}
\end{equation}
showing the mass flux per unit width in the $y$ direction is constant and equal to its upstream value. The dimensional form of \eqref{momy1}, outside boundary layers, can be written
\begin{equation}\label{invmom}
 (u^*v^*)_{x^*} + (w^*v^*)_{z^*} =- 2\Omega(u^*-U),
\end{equation}
Integrating \eqref{invmom} over the flow domain, using the divergence theorem, the vanishing of normal velocity on the upper and lower boundaries, the uniform conditions at large distance  and \eqref{mass}, gives
\begin{equation}
 2UD(V_+ -V_-) = - 2\Omega \int (2UD- UH^*(x^*)) \dx^* = -4\Omega U A ,  
\end{equation}
where $H^*(x^*)$ is the local depth, $A$ is the cross-sectional area of the ridge and $V_-$ and $V_+$ are the values of the cross-flow speed far upstream and far downstream.  The generation of negative relative vertical vorticity as fluid columns are compressed moving over the ridge thus generates a clockwise change in flow direction determined by the difference
\begin{equation}
 V_+ -V_- = -2\Omega A/D,  \label{vdiff}
\end{equation}
which becomes progressively less important the deeper the flow. Since the flow geometry is independent of $y$ any value of $V_-$ can be enforced by an appropriate choice of the upstream pressure gradient $p_x$.

{Sufficiently far upstream of the ridge the velocity field is uniform horizontally and an Ekman layer solution satisfies the full nonlinear system \eqref{maineq} for arbitrary $E$ and $\Ro$, giving the solution
\begin{align}
u^*_0 + \rmi v^*_0 = (U+ \rmi V_-)\{1-\frac{\cosh[(1+\rmi)(2E\delta^2)^{-1/2}(z-\delta)]}
 {\cosh[(1+\rmi)(2E)^{-1/2}]}\}.
 \label{Ekvel}
\end{align}
The non-dimensional shear stress on the lower boundary is thus
\begin{equation}
 \frac{\rmd \phantom{z}}{\rmd z}(u_0 + \rmi v_0) \big|_{z=0} =(2E\delta^2)^{-1/2}(1+\rmi)(1+ \rmi V_-/U)\tanh[(1+\rmi)(2E)^{-1/2}],
\end{equation}
rotated by an angle $\pi/4$ from the outer velocity. For sufficiently small Ekman number the $x$-component of shear stress is negative if $V_-$ exceeds $U$, which it will for any ridge of non-zero cross-sectional area if the rotation is sufficiently rapid 
and the interior flow is symmetric about the ridge. 
The equations are no longer parabolic in the direction of increasing $x$. This difficulty is removed here by taking advantage of the $y$ independence of the flow to set $V_-=0$ far upstream (so $p_x\to0$ as $x\to -\infty$) so the upstream shear stress is always in the $x$ direction. For small Ekman numbers the $x$-component of wall-shear far downstream of the ridge is then, from \eqref{vdiff}, 
\begin{equation}
 (2E\delta^2)^{-1/2}(U+2\Omega A/D) 
\label{ffws}
\end{equation}}
and so increases due to the clockwise rotation of the exterior flow. This rotation also affects the structure of the boundary layer and is discussed in greater detail in \S\ref{S:bl}. The boundary conditions on the flow can thus be written
\begin{align}
 (u,v,w)&=(0,0,0), \qquad\text{on } z = \alpha\; b(x), \label{bc1} \\
 (u_z,v_z,w) &=(0,0,0), \qquad\text{on } z = \delta, \label{symm1} \\
 (u,v,w)&\to(u_0,v_0,0)\big|_{V_-=0}, \qquad\text{as } x\to-\infty, \label{far1}
\end{align}
where the non-dimensional parameter, $\alpha=h/L$, gives the average slope of the ridge and $\delta=D/L$ gives the channel aspect ratio. The flow conditions far downstream, outside the Ekman layers, follow from \eqref{vdiff} as
\begin{equation}
 v\to -2\Omega A/UD, \qquad p_x\to-2\Omega^2 AL/DU^2, \qquad\text{as } x\to\infty. \label{farout}
\end{equation}

\section{Boundary layer analysis, $\Re\gg1$}\label{S:bl}
The continuity equation \eqref{cty} allows the introduction of a streamfunction $\psi$ defined through
\begin{equation}
    u=\psi_z, \qquad w=-\psi_x.
    \label{stream}
\end{equation}
Then the momentum equations become\begin{subequations}
    \begin{align}
    -\frac{\partial(\psi,u)}{\partial(x,z)} - \Rom v &= - p_x+\Rem\nabla^2 u , \label{momx} \\
    -\frac{\partial(\psi,v)}{\partial(x,z)} + \Rom (u-1) &= \Rem\nabla^2 v, \label{momy} \\
    -\frac{\partial(\psi,w)}{\partial(x,z)} \phantom{+ \Rom u} &= - p_z +\Rem\nabla^2 w, \label{momz} 
    \end{align}
\end{subequations}
where $\frac{\partial(f,g)}{\partial(x,y)} := f_x g_y - f_y g_x$. Cross-differentiation of \eqref{momx} and \eqref{momz} gives with \eqref{momy}
\begin{subequations}
    \begin{align}
    -\frac{\partial(\psi,\eta)}{\partial(x,z)} &= \Rom v_z +\Rem\nabla^2 \eta, \\
    -\frac{\partial(\psi,v)}{\partial(x,z)} &= \Rom (\psi_z+1) +\Rem\nabla^2 v, 
    \end{align}
    \label{pair}
\end{subequations}
where $\eta=u_z-w_x=\psi_{xx}+\psi_{zz}=\nabla^2\psi$ is the $y$ component of vorticity. The boundary conditions on $\psi$ and $v$ are
\begin{align}
 (\psi,\psi_z,v) &= (0,0,0), \qquad\text{on } z=\alpha\; b(x), \label{bcbot}\\
 (\psi,\psi_{zz},v) &= (\delta,0,0), \qquad\text{on } z=\delta, \label{bctop}\\
  (\psi,v)&\to(\int u_0\rmd z,v_0), \qquad\text{as } x\to-\infty. \label{bcfar}
\end{align}

\subsection{The outer flow at large Reynolds number}
\label{sec:outer}

In the limit, $Re\to\infty$ with all variables of order unity, \eqref{pair},\eqref{bcbot},\eqref{bcfar} reduce to
\begin{subequations}
    \begin{align}
    -\frac{\partial(\psi,\eta)}{\partial(x,z)} &= \Rom \frac{\partial(x,v)}{\partial(x,z)}, \\
    -\frac{\partial(\psi,v)}{\partial(x,z)} &=  \Rom \frac{\partial(x,z-\psi)}{\partial(x,z)}, 
    \end{align}
    \label{pair2}
\end{subequations}
giving an inviscid outer flow, subject to the inviscid boundary conditions 
\begin{align}
 \psi &= 0, \qquad\text{on } z=\alpha\; b(x), \label{bcbotinv}\\
 \psi &= \delta, \qquad\text{on } z=\delta, \label{bctopinv}\\
 (\psi,v)&\to(z,0), \qquad\text{as } x\to-\infty. \label{bcfarinv}
\end{align}
For unidirectional motion (where $\psi_z=u \neq0$ throughout the flow domain) it is convenient to map the flow domain to a rectangle by taking independent variables $(x,\psi)$ and dependent variable $Z(x,\psi)$.
Multiplying equations \eqref{pair2} by the Jacobian  $|\partial(x,z)/\partial(x,\psi)| = Z_\psi$ gives
\begin{subequations}
    \begin{align}
    -\frac{\partial(\psi,\eta)}{\partial(x,\psi)} &= \Rom \frac{\partial(x,v)}{\partial(x,\psi)}, \\
    -\frac{\partial(\psi,v)}{\partial(x,\psi)} &= \Rom \frac{\partial(x,z-\psi)}{\partial(x,\psi)}, 
    \end{align}
    \label{pair3}
\end{subequations}
i.e.
\begin{subequations}
    \begin{align}
    \eta_x &= \Rom v_\psi, \\
     v_x &= -\Rom Z_\psi + \Rom,
    \end{align}
    \label{pair4}
\end{subequations}
which can be combined to give
\begin{equation}
      Z_{\psi\psi} + \Ro^2\eta_{xx}=0.
    \label{pair5}
\end{equation}
Although the system \eqref{pair4} and equation \eqref{pair5} appear linear, in $(x,\psi)$ coordinates
\begin{align}
   \eta &= \nabla^2 \psi =  (1/2) [ (1 + Z_x^2)/Z_\psi^2]_\psi + (Z_x/Z_\psi)_x, \label{vort}
\end{align}
closing the problem for $Z(x,\psi)$ in $(x,\psi)$ space but rendering the field equation nonlinear. 

The far-field boundary condition \eqref{bcfarinv} becomes  
\begin{equation}
  Z(x,\psi)\to\psi,  \qquad\text{as } |x|\to\infty. \label{Zfar}
\end{equation}

The boundary condition on the ridge becomes simply a condition on the coordinate line $\psi=0$, giving a problem in a rectangular domain with
\begin{subequations}
    \begin{align}
  Z(x,0)&=\alpha\; b(x),   \label{Zbot1} \\
  Z(x,\delta)&=\delta,   \label{Ztop1} \\
  Z(x,\psi)&\to\psi,  \qquad\text{as } |x|\to\infty.   \label{Zfar1}
    \end{align}
\end{subequations}
These equations are equivalent to those derived for stratified, rotating flow by \citeauthor{Jacobs64a}  \cite{Jacobs64a}, although the change in transverse velocity noted here is not discussed there.

In $(x,\psi)$ coordinates the Cartesian velocity components become 
\begin{equation}
  u(x,\psi) = 1/Z_\psi, \qquad w(x,\psi) = Z_x/Z_\psi,
\end{equation}
giving the outer flow velocity component along a streamline as
\begin{equation}
  U_\text{out}(x,\psi) = (u^2+w^2)^{1/2}=(1 + Z_x^2)^{1/2}/Z_\psi. \label{U0def}
\end{equation}
\subsection{The inner, boundary-layer solution}
\label{sec:inner}

For Rossby numbers of order unity, the outer solution matches the no-slip condition on the ridge in a viscous layer of thickness $\Re^{-1/2}$. 
Thus introduce the coordinate $\zeta=\Re^{1/2}\psi$ orthogonal to the surface $\psi=0$ and the scaled velocity component normal to the 
surface $W=\Re^{1/2}u_\text{n}$ for normal velocity component $u_\text{n}$ which vanishes as $\zeta\to\infty$.  Then the boundary layer is governed by the equations
\begin{subequations}
\begin{align}
UU_\xi+WU_\zeta -\Rom V +P_\xi &= U_{\zeta\zeta}, \label{blx} \\
UV_\xi+WV_\zeta + \Rom (1+U) &= V_{\zeta\zeta}, \label{bly} \\
 P_\zeta &= 0, \\
 U_\xi+W_\zeta&=0,
\end{align}
\label{bl}
\end{subequations}
where $\xi$ is the distance along the surface and $U$ the speed in the $\xi$ direction. Choosing the origins of $\xi$ and $x$ to coincide gives
\begin{equation}
    \xi(x) = \int_0^x \sqrt{1+\alpha^2b^{'2}(x')}\;\rmd x'.
    \label{xidef}
\end{equation}
The boundary conditions on system \eqref{bl} are
\begin{subequations}
\begin{align}
[U,V,W] &= [0,0,0],  &\text{on} \quad\zeta=0,  \\
[U,V] &\to [U_e(\xi),V_e(\xi)],   &\text{as} \quad\zeta\to\infty,  \\
[U,V] &= [U_0(\zeta),V_0(\zeta)],   &\text{at} \quad \xi=\xi_0,  \\
 P &\to P_e(\xi),  &\text{as} \quad\zeta\to\infty,
\end{align}
\label{blbcs}
\end{subequations}
where, from \eqref{U0def}, the exterior velocity components are given by $U_e(\xi)=U_\text{out}(x,0)$, $V_e(\xi)=v(x,0)$, $P_e(\xi) = p(x,0)$ for $x=x(\xi)$, obtained by inverting \eqref{xidef} and $U_0(\zeta)$, $V_0(\zeta)$ are the usual Ekman layer velocities, valid for arbitrary $\Re$ 
sufficiently far upstream of the ridge, at some $-\xi_0\gg1$, so, from \eqref{Ekvel},
\begin{equation}\label{Ekvel2}
 U_0(\zeta) +\rmi V_0(\zeta) = U\{1-\exp[-(1+\rmi)(2\Ro)^{-1/2}\zeta]\}.
\end{equation}
For sufficiently low ridges, system \eqref{bl} is parabolic and so can be numerically integrated by stepping downstream from the known upstream profile at $\xi=\xi_0$.

\subsection{The shallow flow limit, $\delta\ll1$}

In general the solution of \eqref{pair5} and \eqref{vort} can be found only numerically. However for shallow flows where $\delta\ll1$, \eqref{pair5} reduces to $Z_{\psi\psi}=0$ with solution
\begin{equation}
 Z(x,\psi) = \alpha b(x) (1-\psi/\delta) + \psi,
\end{equation}
giving the depth-independent horizontal velocity components
\begin{equation}
 u(x,\psi) = [1-\Delta b(x)]^{-1},\quad v(x,\psi) = -(\Delta/\Ro)\int_{-\infty}^x b(x')\rmd x',
 \label{shallowuv} 
\end{equation}
depending solely on the ridge shape and the fractional height $\Delta=\alpha/\delta$. The velocity along the lower boundary is thus, from \eqref{U0def},
\begin{align}
U_e(\xi) &=u(x,0)\sqrt{1+\alpha^2b^{'2}} \nonumber\\
&= \sqrt{1+\alpha^2b^{'2}}/[1-\Delta b(x)], \qquad x=x(\xi).
\end{align}

The shallow flow solution \eqref{shallowuv} provides precisely the outer flow required for the numerical solution of the boundary layer equations \eqref{bl} as formulated in Appendix \ref{S:blnumerics}. 
\begin{figure}
  \begin{center}
    \includegraphics[width=0.48\textwidth]{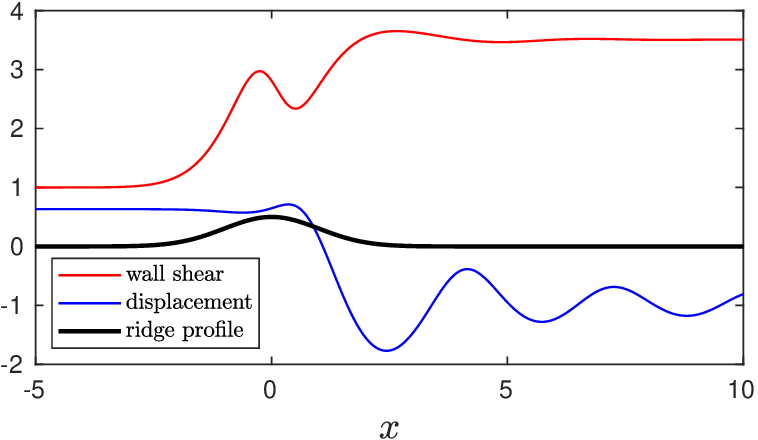}\hfill 
    \includegraphics[width=0.48\textwidth]{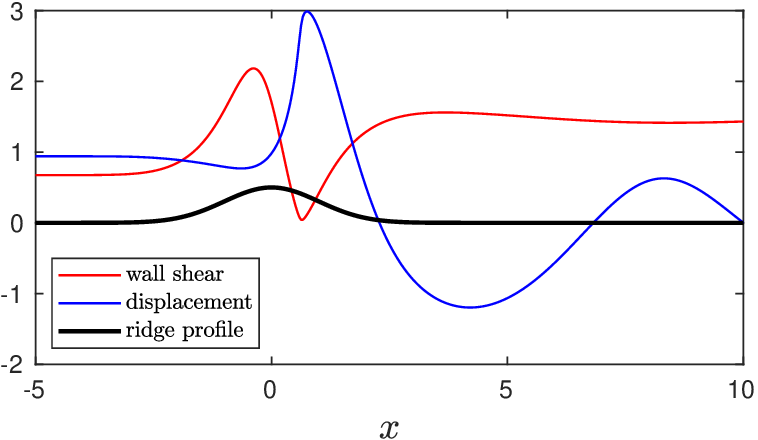}\\
\phantom{.}\hspace{0.20\textwidth} (a) \hspace{0.50\textwidth}  (b) \hfill\phantom{.}\\
  \end{center}
\caption{The shear stress  (red) in the $x$-direction at the wall and the boundary-layer displacement 
thickness (blue) as functions of position $x$ for a Gaussian ridge of maximum fractional depth $\Delta=0.5$. (a) $\Ro=0.5.$ (b) $\Ro=1.1.$ \label{f:wsd}}
\end{figure}
\begin{figure}
  \begin{center}
    \includegraphics[width=0.40\textwidth]{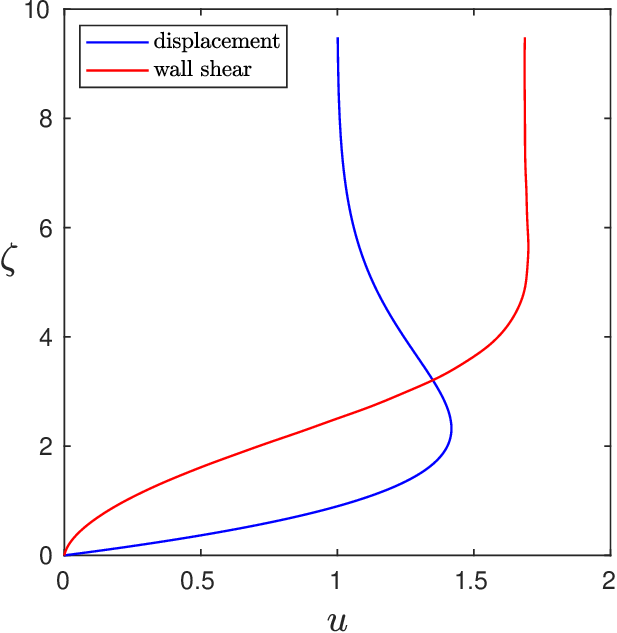}
  \end{center}
\caption{Profiles of $U$, the $x$-component of velocity, as a function of the scaled normal coordinate, $\zeta$ for a Gaussian ridge of maximum fractional depth $\Delta=0.5$ and $\Ro=1.1$, as in figure \ref{f:wsd}(b). The red curve gives the profile at $x=0.63$ corresponding to the minimum in the wall shear and the blue curve gives the profile at $x=4.23$ corresponding to the minimum in the displacement thickness. \label{f:pro_1p1}}
\end{figure}
Two quantities are of particular interest: the scaled boundary-layer displacement thickness
\begin{equation}
 d(x)=\int_0^\infty [1-U(x,\zeta)/U_e(\xi(x))] \;\rmd \zeta,
\end{equation}
and the the magnitude of the shear stress at the wall where the shear stress is given by
\begin{equation}
    \bvec{\tau}(x) = U_\zeta(x(\xi,0)) \bvec{\hat{\xi}} + V_\zeta(x(\xi,0))\bvec{\hat{y}}.
\end{equation}
Figure \ref{f:wsd} shows the first component of ${\tau}(x)$ and  $d(x)$ for a Gaussian ridge $b(x)=\exp(-x^2/2)$ of fractional height $\Delta=0.5$. Negative $\tau_1(x)$ corresponds to recirculation in the planes $y = $ constant.
In figure \ref{f:wsd}(a) rotation is reasonably strong ($\Ro=0.5$). From \eqref{farout} the flow direction thus rotates by $\tan^{-1}[(\Delta/\Ro)\int b(x)\rmd x] = \tan^{-1}\sqrt{2/\pi}$ clockwise on passing over the ridge and the component of the velocity in the $x$-direction increases, as does the shear stress at the wall as in \eqref{ffws}. 
The adverse pressure gradient in the lee of the ridge causes a small local decrease in wall stress. The rotation of the external velocity also leads to an internal maximum of the downstream velocity which exceeds the free-stream velocity (as in figure \ref{f:pro_1p1} for $\Ro=1.1$). This increases the volume flux in the boundary layer compared with that in the free stream and leads to a negative displacement thickness downstream of the ridge.  The adjustment to the new displacement thickness is through a viscously damped oscillation with wavelength determined by a balance between the advection and Coriolis terms in \eqref{blx}, \eqref{bly} and so is of order $\Ro$, as can be seen by comparing figures \ref{f:wsd}(a) with \ref{f:wsd}(b) (for $\Ro=1.1$) where the wavelength is approximately 2.2 times longer.  In figure \ref{f:wsd}(b) nonlinear advection is more 
important making the adverse pressure gradient in the lee of the ridge larger and causing the wall shear stress to fall to a minimum of approximately 0.04. For larger $\Ro$ the wall shear stress becomes negative corresponding to a region of  reversed flow near the wall. The thickening of the boundary layer near the point of minimum wall shear, even for unidirectional flow, is visible in the rapid increase of the displacement thickness there. Figure \ref{f:pro_1p1} shows profiles of $U$ as a function of the scaled normal coordinate, $\zeta$ for the solution of figure \ref{f:wsd}(b). The red curve gives the profile at $x=0.63$ corresponding to the minimum in the wall shear and so is almost vertical at the wall. The blue curve gives the profile at $x=4.23$ corresponding to the minimum in the displacement thickness, showing the jet-like flow in the interior of the layer which increases the mass flux above the corresponding mean flow flux, thinning the layer and giving the negative displacement thickness. 

\section{Numerical simulation of the full equations for shallow flow}
\label{sec:num:shallow}

{We present here  numerical {simulations} of the full nonlinear equations at high $Re$ and small channel aspect ratio $\delta$, comparing them to the boundary layer integrations (for the limit $Re\to \infty, \delta\to 0$) in Section~\ref{S:bl}. The geometry is fixed independently of $\delta$ by the vertical scaling $\hat{z}=z^*/D=z/\delta$. The lower boundary is 
taken as the Gaussian ridge $\hat{z}=\Delta \exp(-x^2/2)$, with $\Delta=0.5$, as in Figs. \ref{f:wsd} and \ref{f:pro_1p1}. }
For the computations, we cut the domain at $x\pm 10$ and define 
\begin{align*}
 \Omega^s := \{ (x,z)\in \mathbb{R}^2, \, -10<x<10, \,b(x)<\hat{z}<1\}.
\end{align*}

We introduce a sub-tesselation ${\cal T}_h$ of $\Omega^s$ into triangles $T\in {\cal T}_h$ and use 
the following finite element spaces of continuous piecewise linear finite elements on ${\cal T}_h$
\begin{align*}
 {\cal V}_h &:= \{ v\in C(\bar \Omega^s)^3, \, v_i|_T \in P_1(T)\, \forall T\in {\cal T}_h,\, \,v_i=0 \text{ on } \Gamma_i^d, \, i=1,2,3\},\\
  {\cal L}_h &:= \{ p\in C(\bar \Omega^s), \, p|_T \in P_1(T)\, \forall T\in {\cal T}_h\}.
\end{align*}
The Dirichlet part of the boundary $\Gamma_i^d$ differs slightly between the components. For $i=1,2$ it 
is given by the lower boundary, where no-slip
conditions \eqref{bc1} are imposed {and the left (upstream) boundary, where the  inflow profile~\eqref{Ekvel2}
is imposed for $U=1$}.
$\Gamma_3^d$ includes additionally
the upper (symmetry) boundary, where the symmetry condition \eqref{symm1} is imposed. On the right downstream boundary the
\textit{do-nothing} conditions
 \begin{align*}
  Re^{-1} \partial_n u - p n =0, \quad \partial_n v = \partial_n w  = 0
 \end{align*}
 are used, where $n$ denotes the exterior normal vector on $\partial\Omega$.
 
 In order to abbreviate notation, we write $U:=(u,v,w), \bar U:=(u,w), \bar\nabla:= (\partial_x,\partial_{\hat{z}})$ and 
 $\bar\nabla_\delta:= (\partial_x,\delta^{-1}\partial_{\hat{z}})$. By $U^d=(u^d, v^d, w^d)$ we denote an extension of the Dirichlet data
 to $\Omega^s$, i.e.$\,$the 
 inflow profile \eqref{Ekvel} for $u$ and $v$ on the upstream boundary and the zero function for $w$
 on the remaining Dirichlet parts.

The discrete variational formulation corresponding to \eqref{momx1}-\eqref{momz1} reads: \textit{Find 
$U\in U^d + {\cal V}_h, p\in {\cal L}_h$, such that}
\begin{eqnarray}
\begin{aligned}\label{VarForm}
 (\bar U&\cdot \bar\nabla U, \phi)_{\Omega^s} + Re^{-1}(\bar \nabla_\delta U, \bar \nabla_\delta \phi)_{\Omega^s} 
 - Ro^{-1} (v,\phi_1)_{\Omega^s} \\ &+ Ro^{-1}(u-1,\phi_2)_{\Omega^s}
 - (p,\bar \nabla \cdot \phi)_{\Omega^s} + (\bar\nabla \cdot\bar U, \eta)_{\Omega^s}\\
 &\qquad+ s_u(U,\phi)  
 + s_p(p,\eta) =0\qquad\forall \phi\in {\cal V}_h, \eta \in {\cal L}_h.
\end{aligned}
\end{eqnarray}
  Due the lack of inf-sup-stability of the discrete spaces and to handle convection-dominated regimes for 
  high $Re$, we add the weakly consistent \textit{Continuous Interior Penalty} stabilisation terms defined by
  \begin{align*}
   s_u(U,\phi) &= \gamma_u \sum_{e\in{\cal E}_h} |e|^2 (\jump{\nabla U}, \jump{\nabla \phi})_e, \\
   s_p(p,\eta) 
   &= \gamma_p \sum_{e\in{\cal E}_h} \max\big\{|e|^3\frac{\delta}{\nu}, \frac{|e|^2}{\|\bar U\|}\big\}
 (\jump{\bar\nabla p}, \jump{\bar\nabla \eta})_e, 
  \end{align*}
  where $\jump{\cdot}$ denotes the jump operator over interior edges $e\in {\cal E}_h$ and $\gamma_u,\gamma_p$ are positive parameters~\cite{Burmanetal2006}.

 
 All the simulation results have been obtained with the finite element library FreeFem++ \cite{FreeFem}. 
 For the visualisation, we use the visualisation tool ParaView \cite{ParaView}.
 We use a triangulation with $192\,682$ 
 vertices ($379\,024$ triangles), which is highly refined along the lower boundary, in order to resolve the boundary layer for high $Re$. To solve 
 the non-linear system of equations, 
 we use a fixed-point iteration based on the Oseen linearisation of the Navier-Stokes equations: For $l=1,\dots$
 \textit{find 
$U^l\in U^d + {\cal V}_h, p^l\in {\cal L}_h$, such that}
\begin{align*}
 \big(\bar\beta^l &\cdot \bar\nabla U^l, \phi\big)_{\Omega^s} + Re^{-1}(\bar \nabla_\delta U^l, \bar \nabla_\delta \phi)_{\Omega^s} 
- Ro^{-1} (v^l,\phi_1)_{\Omega^s} \\
&+ Ro^{-1}(u^l-1,\phi_2)_{\Omega^s}
 - (p^l,\bar \nabla \cdot \phi)_{\Omega^s} + (\bar\nabla \cdot\bar U^l, \eta)_{\Omega^s}\\
 &\qquad+ s_u(U^l,\phi)  + s_p(p^l,\eta) =0\qquad\forall \phi\in {\cal V}_h, \eta \in {\cal L}_h,
\end{align*}
where $\beta^l :=\frac{1}{2}(U^{l-1} + U^{l-2})$ for $l\geq 2$ and $\beta_1=U^0$ for a suitable starting value $U^0$.
Although its rate of convergence is in general worse compared to a standard Newton iteration, we expect this fixed-point 
iteration to be more robust, and 
for this reason better
suitable to give an indication on the existence or non-existence of stationary solutions.
 
 \subsection{The attached boundary-layer regime, $\Ro<1.2$}
 
\begin{figure}
    \begin{center}
    \includegraphics[width=0.49\textwidth]{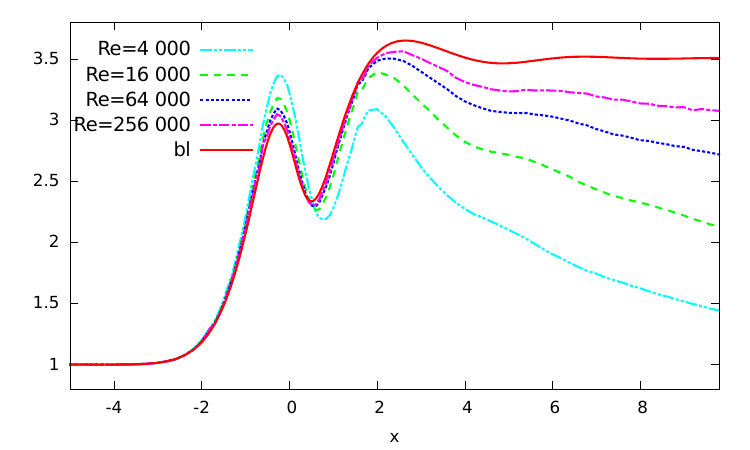}
    \includegraphics[width=0.49\textwidth]{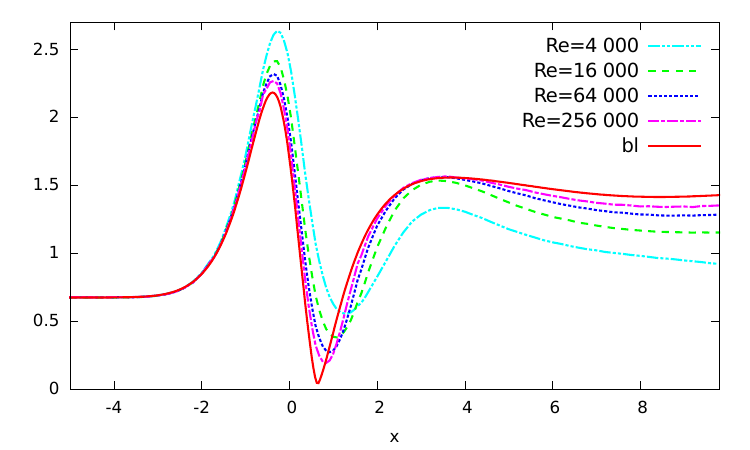}
  \end{center}
  \caption{\label{f:smallD_varRe} The wall shear stress for $\delta=0.1$ and $Ro=0.5$ (left) and $Ro=1.1$ (right) for different $Re$, compared to the boundary layer calculations for $\delta\to 0, Re \to \infty$.}
 \end{figure}

 Figure~\ref{f:smallD_varRe} compares the non-dimensional wall shear stress, 
 \begin{equation*}
\tau = \partial_n u / (E^{1/2} \delta),
 \end{equation*}
 for $\delta=0.1$ and $\Ro\in\{0.5,\,1.1\}$ at increasing Reynolds numbers $\Re = 4000,\, 16000, \,64000\text{ and }256000$ with the boundary layer solutions of Figure \ref{f:wsd},
 corresponding to the limit $\delta\to0$, $\Re\to\infty$. 
  We observe convergence for $Re\to \infty$, although high Reynolds numbers are needed to get close 
 to the results obtained from the boundary layer computations. Notice that we do not expect to fully reach the latter curve, as $\delta$ is fixed to 0.1.  The solution stays attached to the boundary for all tested Reynolds numbers, i.e. up to $Re=1.024\cdot 10^6$. The finite element method appears to give reasonable results up to very large Reynolds numbers.
 
 

 \subsection{Separated, steady, recirculating flow, $Ro\geq 1.2$}
 
\begin{figure}
   \begin{center}
    \includegraphics[width=0.49\textwidth]{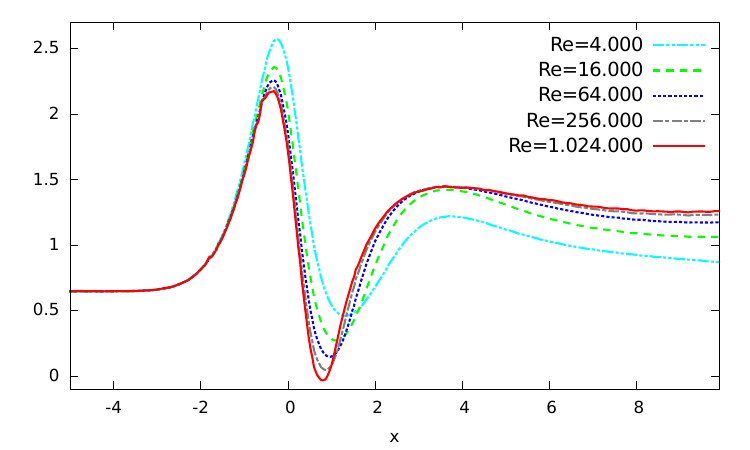}\hfil
    \includegraphics[width=0.49\textwidth]{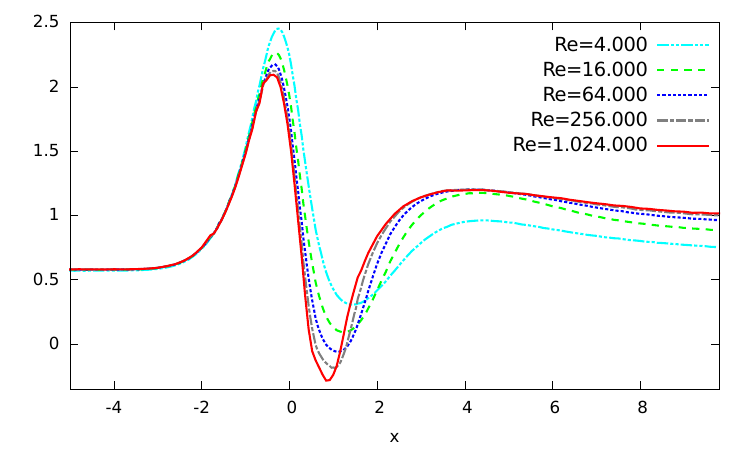}
  \end{center}
 \caption{\label{f:smallD_varRe1215} The wall shear stress for $\delta=0.1$ and $Ro=1.2$ (left) and $Ro=1.5$ (right) for various $Re$.}
 \end{figure}

 For $Ro\geq 1.2$ \eqref{blx}-\eqref{bly} can not be used to compute the inner boundary-layer solution, as the solution detaches in the lee of the ridge (for $\delta\to 0, Re\to \infty$) and the system is no longer parabolic.  The finite-element solution has no such restrictions. Figure~\ref{f:smallD_varRe1215} shows numerical results for fixed $\delta=0.1$ and $Ro=1.2$ and $Ro=1.5$, at various values of $\Re$. For $Ro=1.2$ the boundary layer 
 detaches at $Re=1.024\cdot 10^6$, i.e. the wall shear stress takes negative values in the lee of the ridge, at $x\approx 0.8$. For $Ro=1.5$, this happens earlier, by $Re\geq64\,000$.
 For both Rossby numbers, the method gives stationary states up to (at least) $Re=1.024\cdot 10^6$.
 
\begin{figure}
   \begin{center}
   \begin{subfigure}{0.45\textwidth}
    \includegraphics[width=\textwidth]{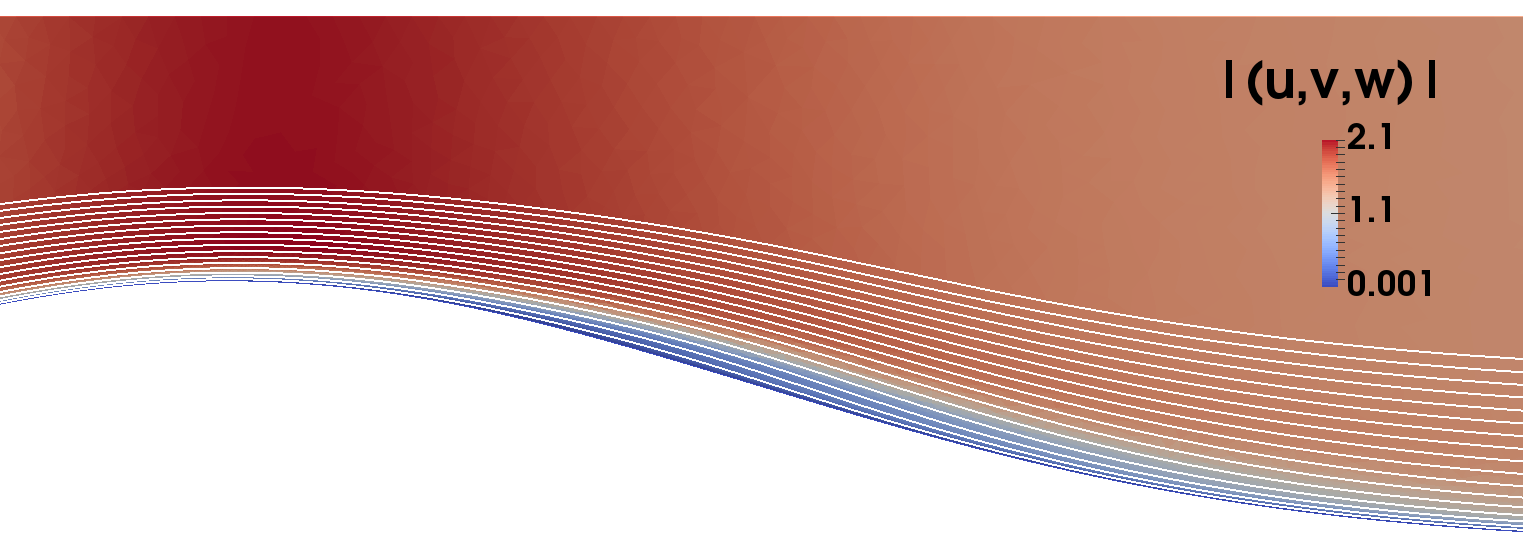}\\
    \includegraphics[width=\textwidth]{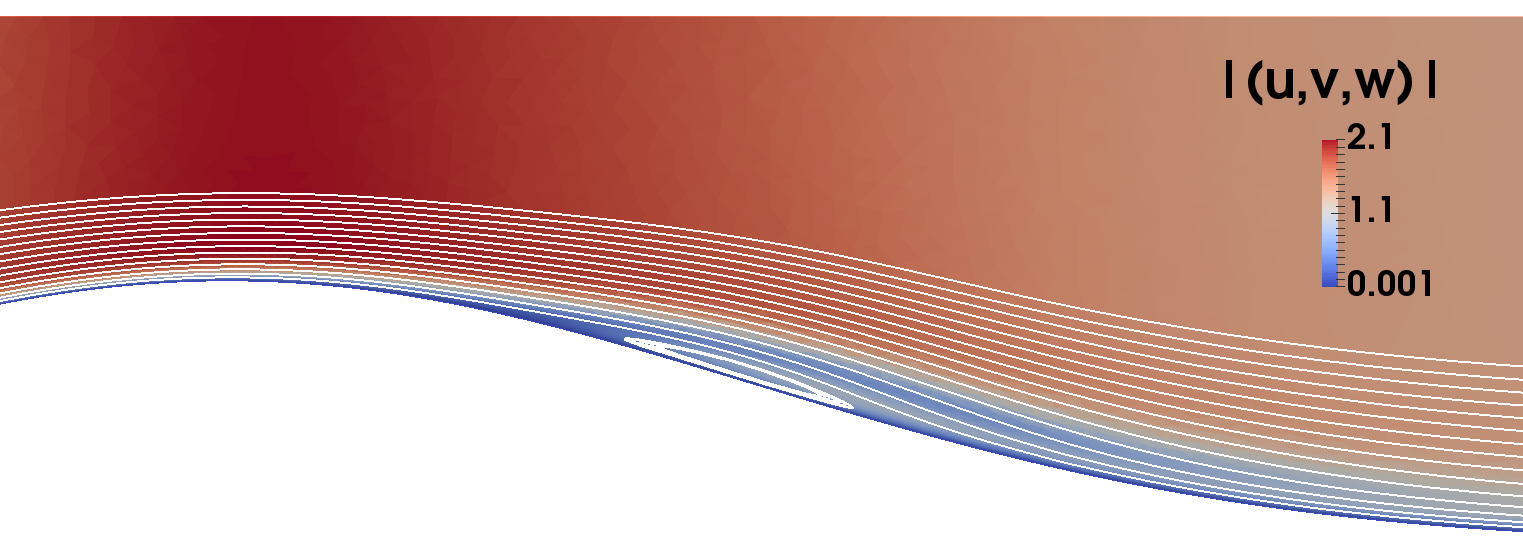}\\
    \includegraphics[width=\textwidth]{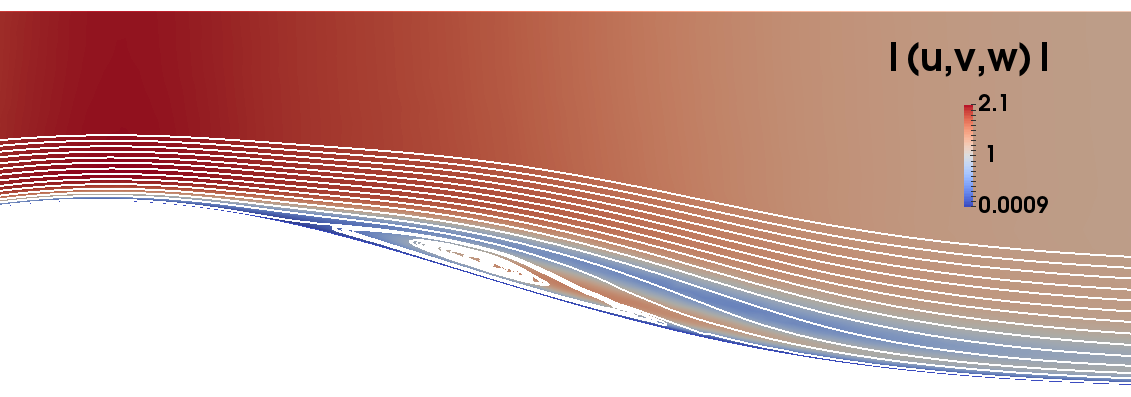}
    \end{subfigure}\hfil
    \begin{subfigure}{0.35\textwidth}
    \includegraphics[width=\textwidth]{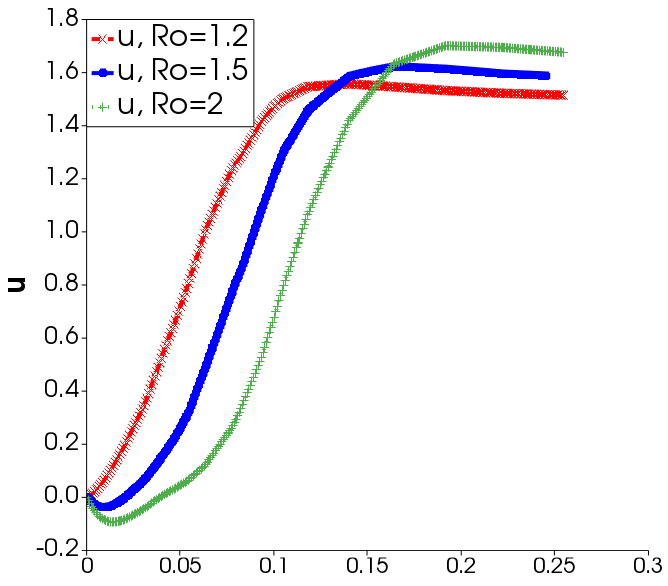}
    \end{subfigure}
  \end{center}
 \caption{\label{f:stream} Left: Streamlines in the lee of the ridge at $Re=256000$ for $Ro=$ 1.2, 1.5 and 2. The flow detaches for $Ro=1.5$ and $Ro=2$ at $x\approx 0.75$ and $x\approx 0.6$, respectively. Right: Profiles  normal to the lower boundary through the center of the eddy of the horizontal velocity $u$ for these cases.}
 \end{figure}

 Figure~\ref{f:stream} (top) shows isolines in the $x-z$ plane of the streamfunction for $Re=256\,000$ and $Ro=1.2$, $Ro=1.5$ and $Ro=2$. Since the $y$-component of velocity is non-zero these are vertical sections of stream surfaces, referred to subsequently, for brevity, as streamlines.   While for $Ro=1.2$ the flow is still attached to the boundary, the flow detaches for $Ro=1.5$ and $Ro=2$ and a separation bubble becomes clearly visible.  Figure~\ref{f:stream} (bottom) shows vertical profiles through the eddy of the horizontal velocity $u$. The negative values show the reverse flow for $Ro=1.5$ and $Ro=2$.
 
\begin{figure}
   \begin{center}
    \includegraphics[width=0.48\textwidth]{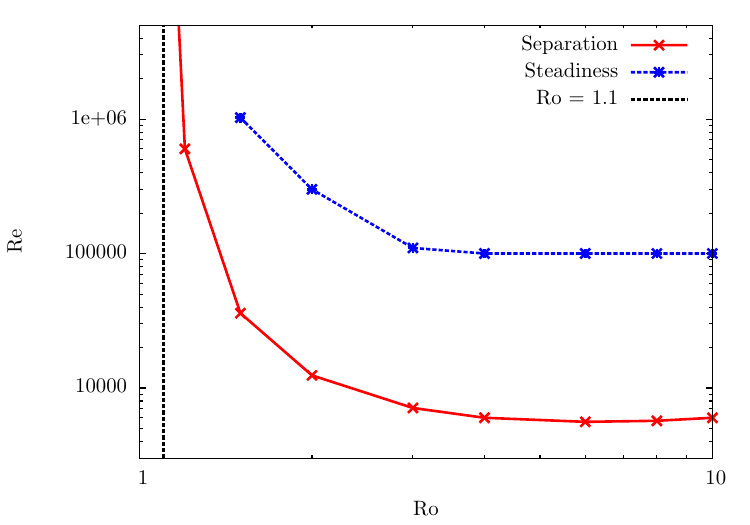}
  \end{center}
 \caption{\label{fig.sepshallow}Shallow flow, $\delta=0.1$. Red solid curve: The smallest value of $Re$, at fixed $Ro$, for which the flow separates. For all pairs $(Ro, Re)$ above the line a separation bubble occurs in the lee of the ridge. Blue dotted curve: The largest value of $Re$, at fixed $Ro$, for which a stationary solution was obtained by  the fixed point algorithm. The vertical dashed line shows the upper bound at $Ro=1.1$ for attached flow in the boundary layer analysis of \S\ref{S:bl}.}
 \end{figure}

 Figure~\ref{fig.sepshallow} shows the critical Reynolds numbers (red solid curve) for boundary layer separation as a function of Rossby number.  Separation occurs at all points above the curve while at points below the curve the flow remains attached. For the ridge geometry here, the flow remains attached for Reynolds numbers, $Re\leq 5\,600$, at all rotation rates, with separation depending solely on the Reynolds number for $\,Ro\geq 6$. At smaller Rossby numbers, $Ro< 1.2$, when rotational effects become significant, the flow remained attached to the boundary for all tested configurations, i.e.$\,$up to at  least $Re=1.024\cdot 10^6$. With decreasing $Ro$ the curve approaches the vertical dashed line showing the upper bound at $Ro=1.1$ for attached flow in the boundary layer analysis of \S\ref{S:bl}.
 At any fixed $Ro>1.1$ there exists a critical Reynolds number, higher than the Reynolds number for separation, above which the fixed-point algorithm is unable to find a stationary solution. This gives an indication
 that above some value of $Re$ no stable stationary solution exists. The values for the disappearance of stationary solutions are shown by the blue dotted curve in Figure~\ref{fig.sepshallow} which shows the same qualitative behaviour as the separation Reynolds number curve. For Rossby numbers greater than 1.2 there is thus a considerable range of Reynolds numbers where a steady recirculating separation bubble is present in the lee of the ridge.

\section{Deep flow over a ridge}
\label{sec:num:deep}

The shallow domain of \S\ref{sec:num:shallow} suppresses inertial waves. To allow inertial waves and assess the effect of the absence of an upper rigid boundary we cut the computational domain at $z=100$ and impose the boundary conditions $(u,v,w) = (1,0,0)$ at the left inflow boundary and on the top boundary. The ridge profile is taken as $z=\alpha\exp(-x^2/2)$, with $\alpha=1/2$, giving ridge slopes that are 10 times those of \S\ref{sec:num:shallow}. This is reflected below in separation at much smaller values of the Reynolds number. The variational formulation is given by \eqref{VarForm} for $\delta=1$, where the space ${\cal V}_h$ is modified to accommodate for the Dirichlet boundary conditions on the upper boundary for all variables.

\begin{figure}
\centering
\begin{minipage}{0.35\textwidth}
 \includegraphics[width=\textwidth]{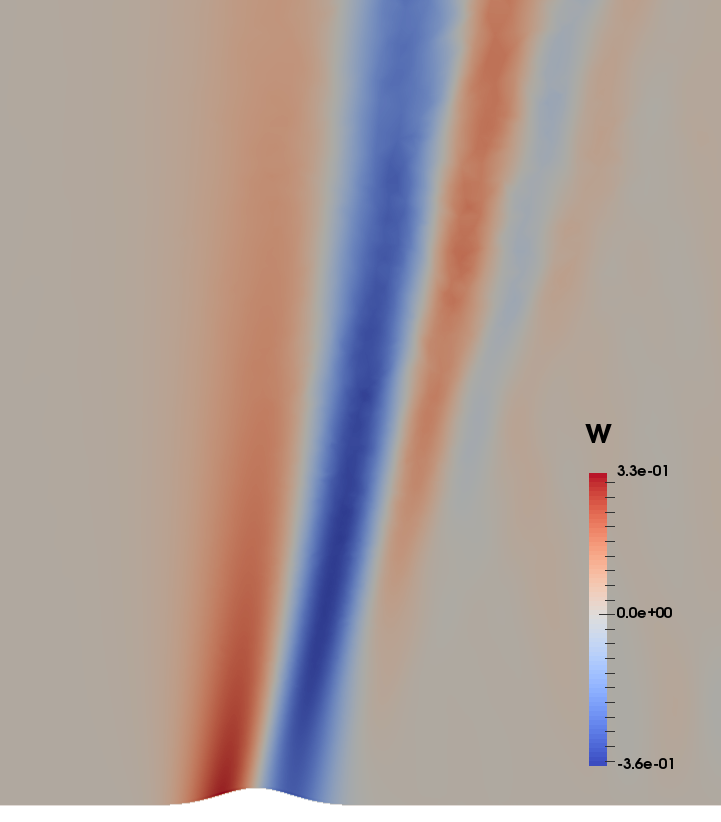}
\end{minipage}\hfil
\begin{minipage}{0.35\textwidth}
 \includegraphics[width=\textwidth]{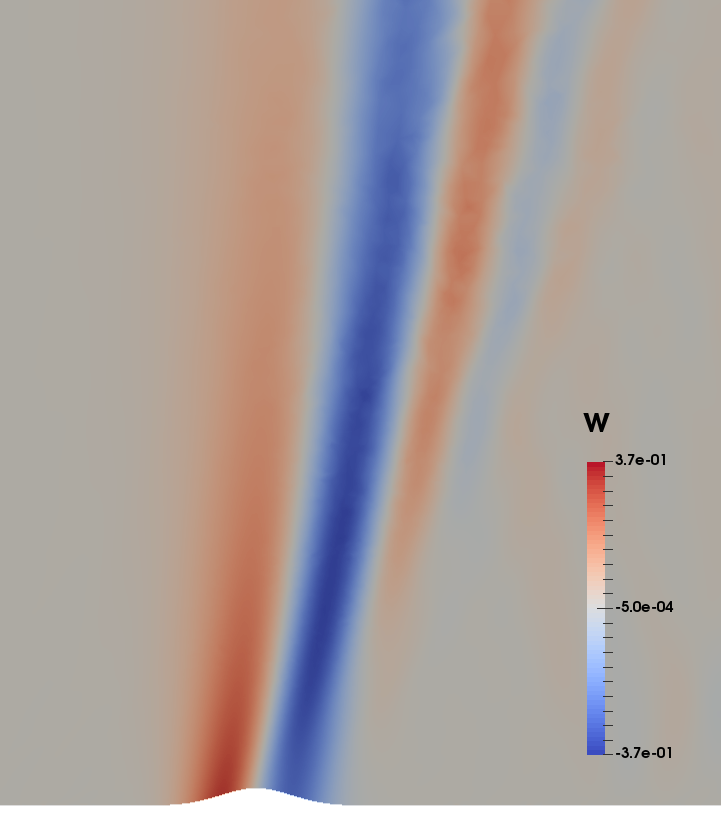}
\end{minipage}
%
 \caption{\label{fig.bump} The vertical velocity component $w$ for deep flow with Ro=0.1 at Re=400 (left) and Re=10\,000 (right). 
 The flow field is almost unaltered when the Reynolds changes by a factor of 25. The mesh is highly refined around the ridge and the figure shows approximately half the simulation domain.}
\end{figure}

Figure~\ref{fig.bump} shows the vertical velocity component $w$  for $Ro=0.1$ at $Re=400$ and $10\,000$. The flow patterns are very similar. This accords with the analysis in \cite{Johnson82a, Machicoane18} showing that the leading order viscous effect in deep flow at small Rossby number is the introduction of a downstream decay of the lee wave wake with a scale of order $Re^{-1}$ and a downstream displacement of the phase lines at large $Re$ of order $(RoRe)^{-1}$.
 
\begin{figure}
   \begin{center}
    \includegraphics[width=0.48\textwidth]{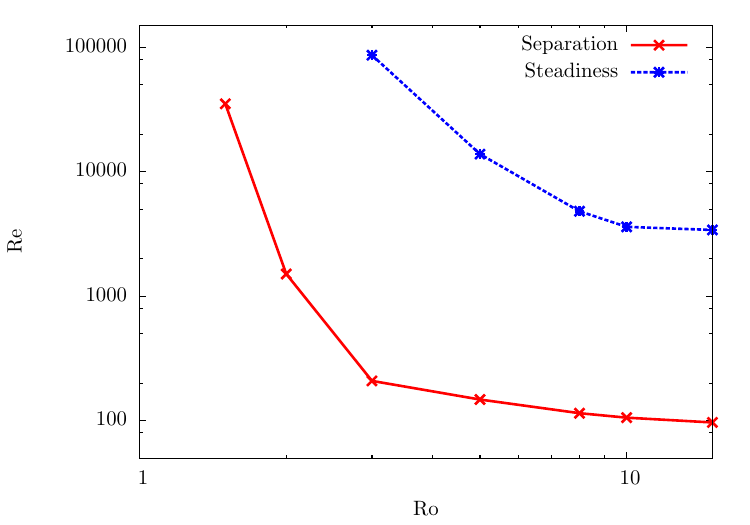}
  \end{center}
 \caption{\label{fig.sepconvdeep} As figure~\ref{fig.sepshallow} but for deep flow. Red solid curve: The smallest value of $Re$, at fixed $Ro$, for which the flow separates. Blue dotted curve: The largest value of $Re$, at fixed $Ro$, for which a stationary solution was obtained by 
 the fixed point algorithm.}
 \end{figure}

Figure~\ref{fig.sepconvdeep}, as for figure~\ref{fig.sepshallow}, shows at fixed Rossby number the largest values of Reynolds number which the flow remained attached (red curve) and at which the fixed point iteration converged (blue curve). The qualitative behaviour of the curves is very similar but the critical Reynolds numbers are an order of magnitude smaller in the deep flow. As noted above, this is a consequence of the ridge slope's being an order of magnitude larger here.
At fixed Rossby number, greater than 1.5, there again exists  a considerable range of Reynolds numbers where a steady recirculating separation bubble is present in the lee of the ridge.

\begin{figure}
\begin{center}
\begin{overpic}[width=0.45\textwidth]{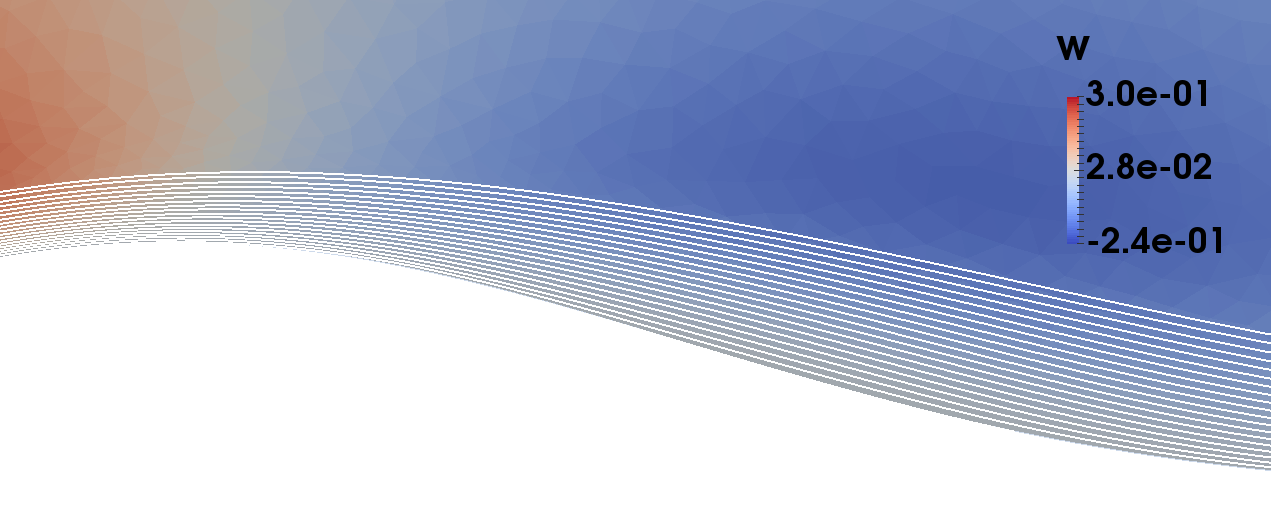}\hfil
\put(0,5){(a)}
\end{overpic}\hfil
\begin{overpic}[width=0.45\textwidth]{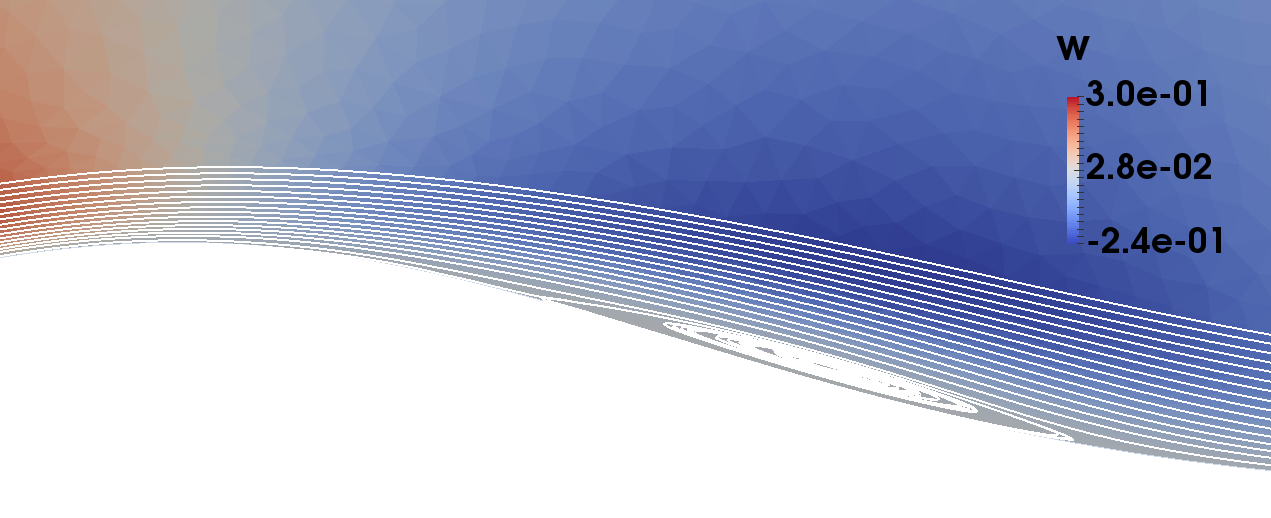}
\put(0,5){(b)}
\end{overpic}
\end{center}
 \caption{\label{fig.sep} The vertical velocity $w$ and streamlines in the lee of the ridge for deep flow and weak rotation,  $\Ro=3$. (a) $\Re=200$, attached flow, (b): $\Re=1000$, separated flow with a stable separation bubble.}
\end{figure}

Figure~\ref{fig.sep} shows the vertical velocity $w$ and streamlines in the lee of the ridge for deep flow and weak rotation with $Ro=3$ at $Re=200, 1000$. The flow is attached for $Re=200$ but has separated by $Re=1000$ with a steady recirculating separation bubble visible.

\section{Flow over a horizontal cylinder}
\label{sec:num:cyl}

To capture more closely the geometry of the experiments reported in \cite{Machicoane18}, this section considers 
flow over a horizontal right circular cylinder. The variational formulation is again given by \eqref{VarForm} with $\delta=1$. We use an ellipsoidal simulation domain with semi-axes of size 60 (vertically) and 150 (horizontally) from which a circle of radius $0.5$ (representing an infinitely long cylinder in transversal direction), is extracted. We set $U^d=(1,0,0)$ on the outer boundary of the ellipsoid and $U^d = (0,0,0)$ on the the cylinder.

\begin{figure}
\begin{center}
\begin{overpic}[width=0.40\textwidth]{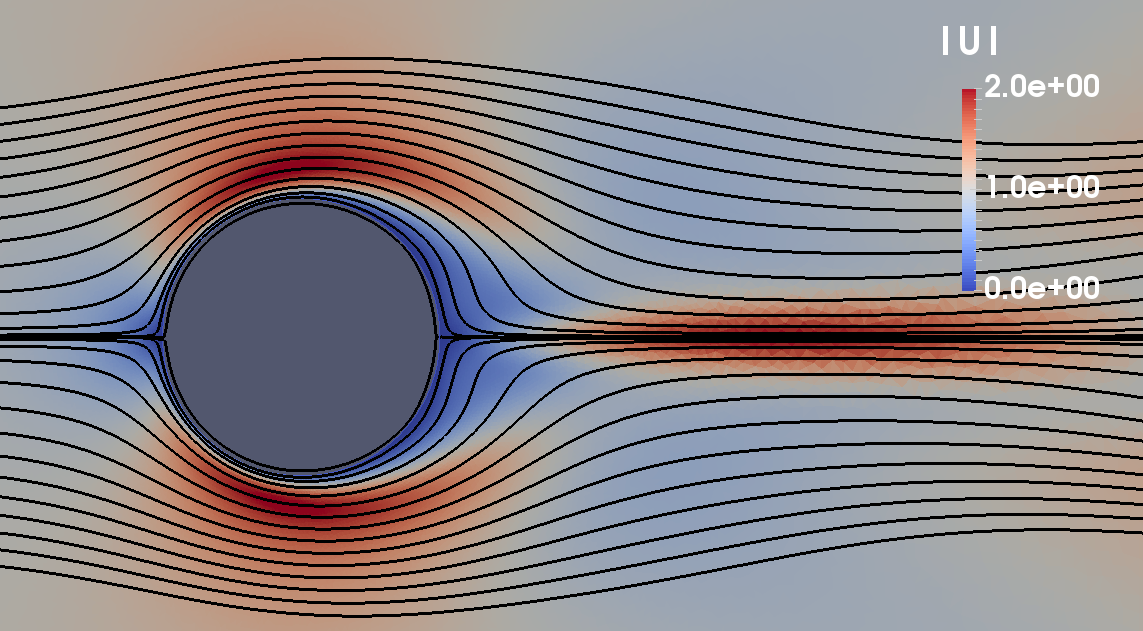}
\put(0,1){(a)}
\end{overpic}\hfil
\begin{overpic}[width=0.40\textwidth]{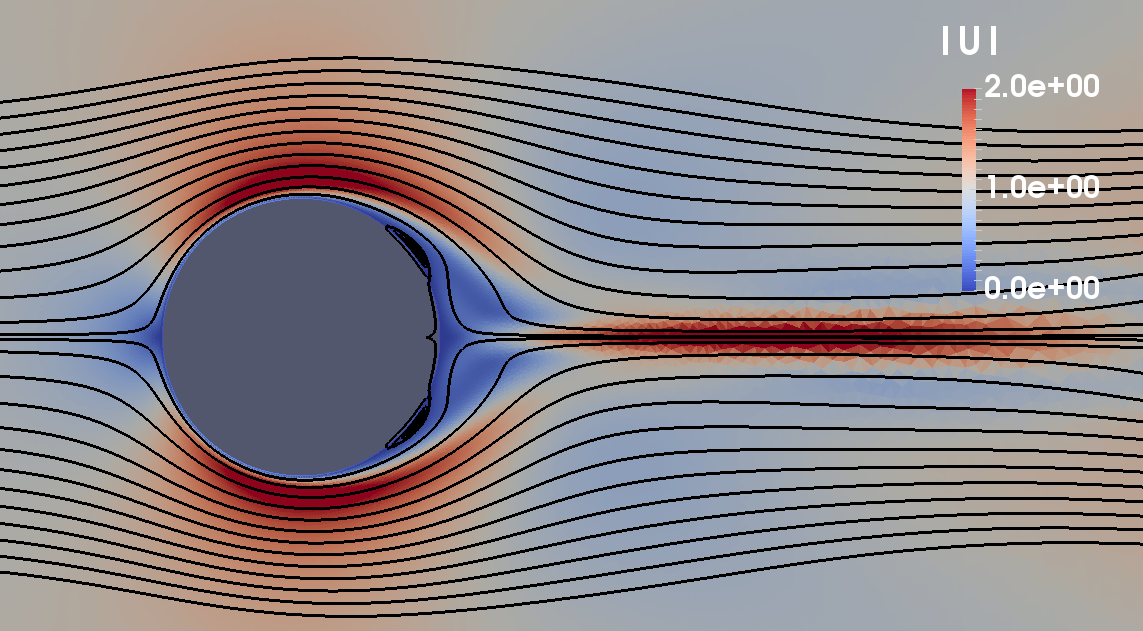}
\put(0,1){(b)}
\end{overpic}\\[0.1cm]
\begin{overpic}[width=0.40\textwidth]{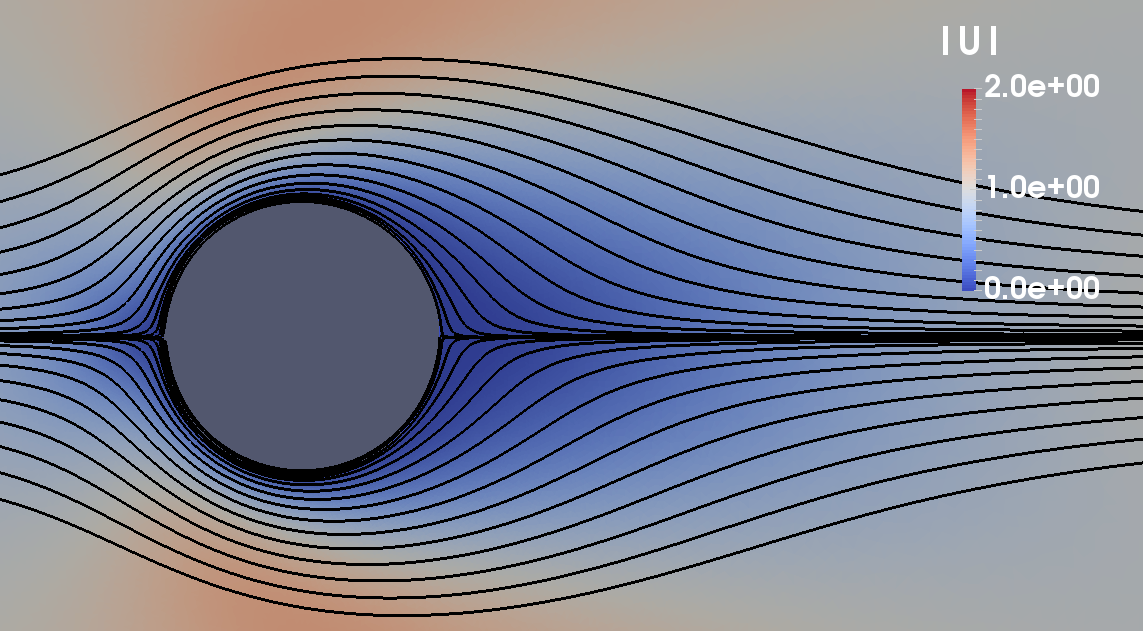}
\put(0,1){(c)}
\end{overpic}\hfil
\begin{overpic}[width=0.40\textwidth]{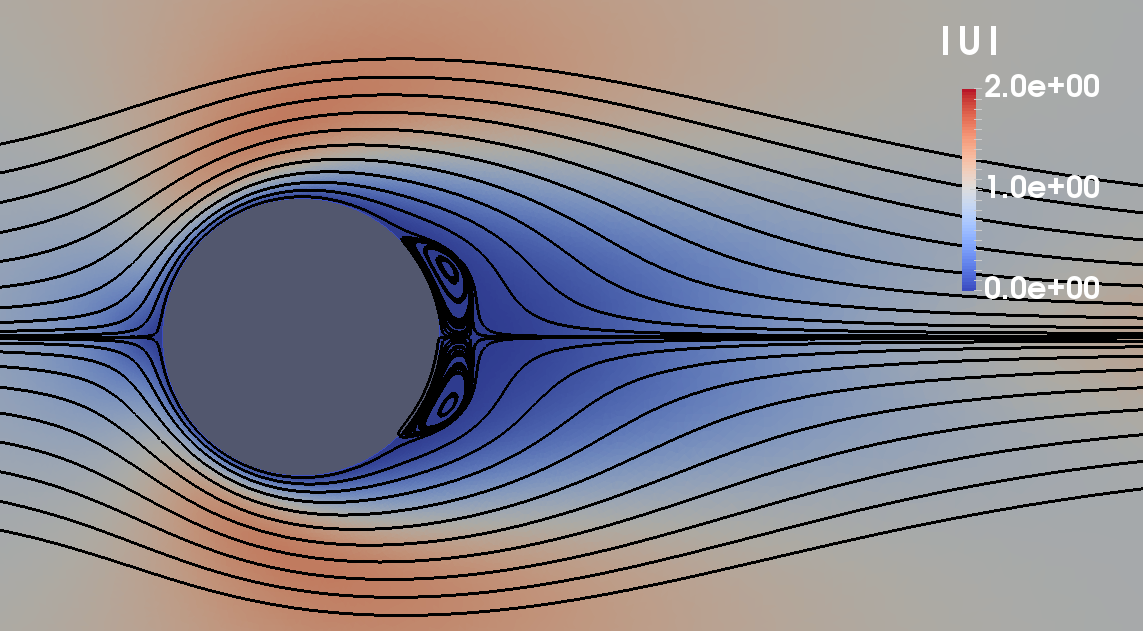}
\put(0,1){(d)}
\end{overpic}
\end{center}
 \caption{\label{fig.Ro001Re1000} Streamlines for $(\Ro,\Re)=$ (a) (0.5,200), (b) (0.5,600), (c) (2,15) and (d) (2,50). The  colouring shows the magnitude of the velocity $(u,w)$. The figure shows only a small part of the simulation domain.}
\end{figure}

\begin{figure}
\begin{center}
    \begin{overpic}[width=0.4\textwidth]{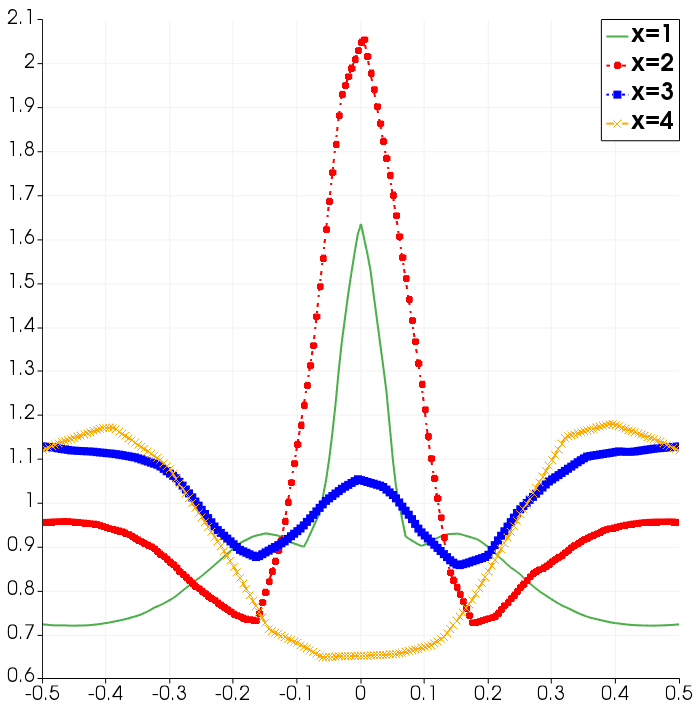}
    \put(-3,50){u}
    \put(50,-2){y}
    \end{overpic}
\end{center}
 \caption{\label{fig.jet} Vertical profiles through the lee jet of the horizontal velocity $u$ at downstream stations $x=1,2,3$ and $4$ for the flow in Fig. \ref{fig.Ro001Re1000}(b) where $\Ro=0.5$ and $\Re=600$.}
\end{figure}

Figure~\ref{fig.Ro001Re1000} shows a pair of streamlines for each of (a,b) $Ro=0.5$ and (c,d) $Ro=2$ at Reynolds numbers $Re$ chosen so the flow is attached to the cylinder at the lower $Re$ (a,c) and detached at the higher (b,d). The lower $Ro$ flow remains attached at significantly higher $Re$ but in both cases the steep obstacle gradient at the rear of the cylinder, and the concomitant strong adverse pressure gradient, induces separation at far lower Reynolds numbers than in flow over the smooth ridge. The inertial wave wake alternatively compresses and accelerates and then expands and decelerates a jet-like flow along the axis in the lee of the cylinder. Figure~\ref{fig.jet} shows vertical profiles through the jet of the horizontal velocity $u$ at downstream stations $x=1,2,3$ and $4$ for the flow in Fig. \ref{fig.Ro001Re1000}(b). The jet is narrow with almost twice the free-stream speed at $x=2$ but wide with a velocity deficit giving a speed of two-thirds the free stream speed at $x=4$.

\begin{figure}
    \centering
    \begin{subfigure}{0.45\textwidth}
    \includegraphics[width=\textwidth]{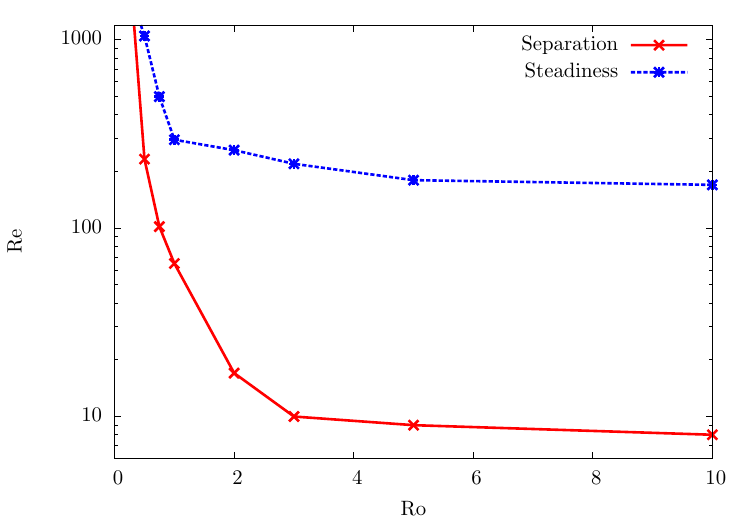}
    \end{subfigure}\\
    \begin{subfigure}{0.7\textwidth}
    \begin{tabular}{l|cccccccc}\small
    $\Ro$ & 0.3 & 0.5 & 0.75 & 1 & 2 & 3 & 5 & 10\\
    $\Re_{\text{crit}}$ &1640 &1050 &500 &295 &260 &220 &180 &170 \\
    \hline
    $\Ro\cdot \Re_{\text{crit}}$ & 492 & 525 & 375 &295 & 520 & 660 & 900 & 1700\\
 \end{tabular}
 \end{subfigure}
     \caption{\label{fig.convcyl} \textit{Top}: As figures~\ref{fig.sepshallow}, \ref{fig.sepconvdeep}, but for deep flow over a cylinder. Red solid curve: The smallest value of $Re$, at fixed $Ro$, for which the flow separates.  Blue dotted curve: The largest value of $Re$, at fixed $Ro$, for which a stationary solution was obtained by 
 the fixed point algorithm.
 \textit{Bottom:} The largest Reynolds number $Re_{\text{crit}}$, for which a stationary solution was obtained by 
 the fixed point algorithm for fixed $Ro$ and product $Ro\cdot Re_{\text{crit}}$. 
 }
\end{figure}

\begin{figure}
   \begin{center}
    \includegraphics[width=0.48\textwidth]{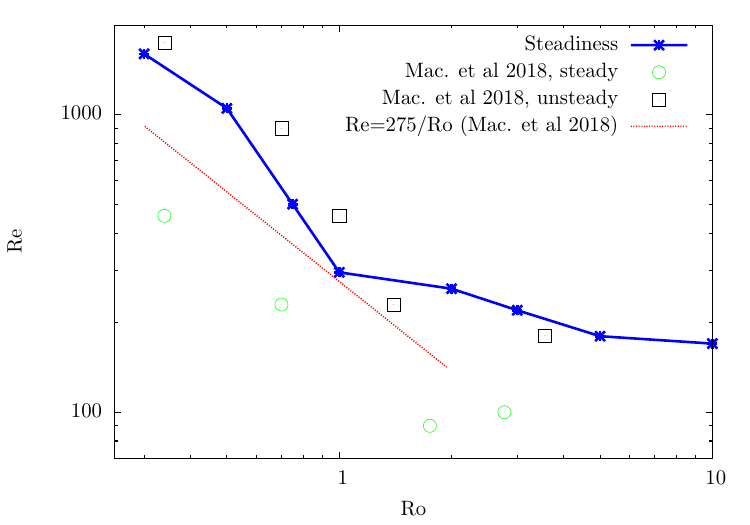}
  \end{center}
 \caption{\label{fig.cyl_comp} Comparison of the critical Reynolds numbers found by numerical simulations (blue curve) against the experimental results in~\cite{Machicoane18}.
 Green circles indicate combinations of $Ro$ and $Re$, for which steady solutions were observed experimentally; black squares indicate unsteady solutions.
 }
 \end{figure}

As figures~\ref{fig.sepshallow} and  \ref{fig.sepconvdeep}, figure~\ref{fig.convcyl} shows for fixed Rossby number the lowest $Re$ for which separation occurs and the highest $Re$ for which the fixed point iteration converged.   The curves behave qualitatively very similarly to those of Figures~\ref{fig.sepshallow} and \ref{fig.sepconvdeep}: as $Ro\to 0$ the critical Reynolds number for convergence, $Re_{\text{crit}}$ increases rapidly, showing the existence of steady solutions at high $Re$, and for weakly rotating flow, $Ro\to\infty$,  $Re_{\text{crit}}$ asymptotes to the constant value ($Re_{\text{crit}} \approx 160$ here) of non-rotating flow. For Reynolds numbers of order 200 and larger the range of values of Rossby numbers for which steady separated solutions exist is small and becomes smaller with increasing Reynolds number, almost vanishing by $Re=1000$. These Reynolds numbers include the parameter regime investigated in \cite{Machicoane18} and supports their observation that separation and unsteadiness appear simultaneously. 
Figure~\ref{fig.cyl_comp}, shows the $Re_{\text{crit}}$ curve over a smaller range of Reynolds numbers with superposed the experimental observations of \cite{Machicoane18} (Figure 7) that are closest to the transition between steady and unsteady solutions. The qualitative form of the numerical curve appears to capture the experimentally observed separation between steady and unsteady flow.  Two experimental points for unsteady flow lie just below the numerical curve. The deviation is within the range of experimental fluctuations and numerical error. In the numerical solution, numerical diffusion could lead to convergence to stationary states on the discrete level that are not stationary states on the continuous level so the critical curve in the continuous limit would lie slightly below the plotted curve. In the experiments even small disturbances to the flow might be of sufficient magnitude to trigger instability below the theoretical threshold. Even with this deviation the numerical curve seems to capture the experimental results well over the whole range. Based on their analysis of the lee-wave wake at large Reynolds number \cite{Machicoane18} postulate that the division between steady and unsteady flow is of the form $Ro\cdot Re_{\text{crit}} = C$ with the experimental results giving $C=275$. This curve is included in Figure~\ref{fig.cyl_comp}
and captures the division in the flow regime there well. The table in Figure~\ref{fig.convcyl} shows that the relation may not capture the division so well in other flow regimes.



\section{Conclusion}
\label{sec:concl}
Motivated by the geophysical relevance of rotating flow over topography in the atmosphere \cite{MasonS79} and ocean and by recent theoretical results and laboratory experiments \cite{Machicoane18} we have discussed steady attached flow, boundary layer separation, transition to unsteady flow and flow patterns in the three numerical examples of shallow and deep flow over a Gaussian ridge and deep flow over a horizontal cylinder using a stabilised equal-order finite element method with Continuous Interior Penalty Stabilisation. The advantage of this technique is that it can obtained steady solutions at high Reynolds numbers. For shallow flow (with aspect ratio $\delta=D/L\to0$) the solutions were compared with integration of the boundary layer equations valid in the limit $Re\to\infty$. The finite-element results for $\delta=0.1$ approached the boundary layer results with increasing Reynolds number and in the region of attached flow the finite-element results converged for Reynolds numbers of order $10^6$. In deep flow over the ridge the inertial wave wake discussed in \cite{Johnson82a, Machicoane18} appeared. The computations for deep flow over a cylinder included the parameter regime of the experiments of \citeauthor{Machicoane18}\cite{Machicoane18} and reproduced many of the features observed there.

In all three cases the flow was steady and attached for sufficiently small Reynolds number. As the Reynolds number was increased at fixed Rossby number the flow separated in the lee of the ridge and a steady separation bubble appeared. At higher Reynolds numbers the fixed point iteration failed to converge. In flow over a cylinder this coincided closely with the appearance of unsteady flow in the experiments of \cite{Machicoane18}. The  Reynolds numbers above which separation and, subsequently, unsteadiness appear decrease by an order of magnitude in moving from shallow to deep flow, as the ridge slope increases by a factor of 10, and by another order of magnitude in moving from the smooth ridge to the cylinder, with its infinite slope at the rear stagnation point. In all cases it appears that sufficiently strong rotation allows the flow to remain attached at arbitrarily large Reynolds numbers.

\appendix

\section{Boundary-layer integration \label{S:blnumerics}}
The method here follows Prandtl (1938), modified to allow expansion in Laguerre polynomials.  In \eqref{bl} write $(U,V)=(U_e+\ut,V_e+\vt)$ so $(\ut,\vt)\to0$ as $\zeta\to\infty$. Then the $\xi$-momentum equation \eqref{blx} becomes
\begin{equation}
 U\ut_\xi+W\ut_\zeta = \ut_{\zeta\zeta} + a\vt - U_{e\xi}\ut, \label{blxt}
\end{equation}
where the far-field pressure satisfies
\begin{equation}
 P_{e\xi} = -U_eU_{e\xi}+aV_e.
\end{equation}
Substituting \eqref{blxt} into the continuity equation and rearranging gives
\begin{equation}
(W/U)_\zeta = -(U_eU_{e\xi} + a\vt + \ut_{\zeta\zeta} )/U^2. \label{blw}
\end{equation}
On the solid boundary, $\zeta=0$,  $U=0$ for all $\xi$, so $U_\xi=0$  and hence $W_\zeta=0$. Thus
\begin{equation}
 \lim_{\zeta\to0}(W/U)= \lim_{\zeta\to0}(W_\zeta/U_\zeta) =0,
\end{equation}
and \eqref{blw} can be integrated from zero on the wall to give $W/U$ at fixed $\xi$. The momentum equations \eqref{blx} and \eqref{bly} then become
\begin{subequations}
\label{blt}
\begin{align}
\ut_\xi &= -(W/U)\ut_\zeta + (\ut_{\zeta\zeta} + a\vt - U_{e\xi}\ut)/U, \\
\vt_\xi &= -(W/U)\vt_\zeta + (\vt_{\zeta\zeta} - a\ut - V_{e\xi}\ut)/U.
\end{align}
\end{subequations}
It is convenient to obtain the outer flow and to discuss the solutions in the unstretched coordinate $x$. Multiplying \eqref{blt} by $\xi_x =\sqrt{1+\alpha^2b^{'2}(x')} $ gives
\begin{subequations}
\label{bltx}
\begin{align}
\ut_x &= -\xi_x[(W/U)\ut_\zeta +  (\ut_{\zeta\zeta} + a\vt)/U] - U_{ex}\ut/U, \\
\vt_x &= -\xi_x[(W/U)\vt_\zeta +  (\vt_{\zeta\zeta} - a\ut)/U] - V_{ex}\ut/U.
\end{align}
\end{subequations}

Equations \eqref{bltx} and \eqref{blw} are solved spectrally in $\zeta$ by expressing $\ut$ and $\vt$ as the product of a decaying exponential 
and a series of Laguerre polynomials through their values at the Laguerre interpolation points, expressing  \eqref{blw} as a first order 
differential equation using a differentiation matrix, inverting to obtain $W/U$ at the interpolation points, and advancing in $x$ using \eqref{bltx} and a stiff ordinary differential equation solver. 


\begin{thebibliography}{15}%
\makeatletter
\providecommand \@ifxundefined [1]{%
 \@ifx{#1\undefined}
}%
\providecommand \@ifnum [1]{%
 \ifnum #1\expandafter \@firstoftwo
 \else \expandafter \@secondoftwo
 \fi
}%
\providecommand \@ifx [1]{%
 \ifx #1\expandafter \@firstoftwo
 \else \expandafter \@secondoftwo
 \fi
}%
\providecommand \natexlab [1]{#1}%
\providecommand \enquote  [1]{``#1''}%
\providecommand \bibnamefont  [1]{#1}%
\providecommand \bibfnamefont [1]{#1}%
\providecommand \citenamefont [1]{#1}%
\providecommand \href@noop [0]{\@secondoftwo}%
\providecommand \href [0]{\begingroup \@sanitize@url \@href}%
\providecommand \@href[1]{\@@startlink{#1}\@@href}%
\providecommand \@@href[1]{\endgroup#1\@@endlink}%
\providecommand \@sanitize@url [0]{\catcode `\\12\catcode `\$12\catcode
  `\&12\catcode `\#12\catcode `\^12\catcode `\_12\catcode `\%12\relax}%
\providecommand \@@startlink[1]{}%
\providecommand \@@endlink[0]{}%
\providecommand \url  [0]{\begingroup\@sanitize@url \@url }%
\providecommand \@url [1]{\endgroup\@href {#1}{\urlprefix }}%
\providecommand \urlprefix  [0]{URL }%
\providecommand \Eprint [0]{\href }%
\providecommand \doibase [0]{https://doi.org/}%
\providecommand \selectlanguage [0]{\@gobble}%
\providecommand \bibinfo  [0]{\@secondoftwo}%
\providecommand \bibfield  [0]{\@secondoftwo}%
\providecommand \translation [1]{[#1]}%
\providecommand \BibitemOpen [0]{}%
\providecommand \bibitemStop [0]{}%
\providecommand \bibitemNoStop [0]{.\EOS\space}%
\providecommand \EOS [0]{\spacefactor3000\relax}%
\providecommand \BibitemShut  [1]{\csname bibitem#1\endcsname}%
\let\auto@bib@innerbib\@empty
\bibitem [{\citenamefont {Mason}\ and\ \citenamefont {Sykes}(1979)}]{MasonS79}%
  \BibitemOpen
  \bibfield  {author} {\bibinfo {author} {\bibfnamefont {P.~J.}\ \bibnamefont
  {Mason}}\ and\ \bibinfo {author} {\bibfnamefont {R.~I.}\ \bibnamefont
  {Sykes}},\ }\bibfield  {title} {\bibinfo {title} {Separation effects in
  {E}kman layer flow over ridges},\ } {\bibfield  {journal} {\bibinfo
  {journal} {Quart. J. Roy. Met. Soc.}\ }\textbf {\bibinfo {volume} {105}},\
  \bibinfo {pages} {129} (\bibinfo {year} {1979})}\BibitemShut {NoStop}%
\bibitem [{\citenamefont {Richards}\ \emph {et~al.}(1992)\citenamefont
  {Richards}, \citenamefont {Smeed}, \citenamefont {Hopfinger},\ and\
  \citenamefont {D'Hi\'eres}}]{RichardsSHCH92}%
  \BibitemOpen
  \bibfield  {author} {\bibinfo {author} {\bibfnamefont {K.~J.}\ \bibnamefont
  {Richards}}, \bibinfo {author} {\bibfnamefont {D.~A.}\ \bibnamefont {Smeed}},
  \bibinfo {author} {\bibfnamefont {E.~J.}\ \bibnamefont {Hopfinger}},\ and\
  \bibinfo {author} {\bibfnamefont {G.~C.}\ \bibnamefont {D'Hi\'eres}},\
  }\bibfield  {title} {\bibinfo {title} {Boundary-layer separation of rotating
  flows past surface-mounted obstacles},\ } {\bibfield  {journal} {\bibinfo
  {journal} {J. Fluid Mech.}\ }\textbf {\bibinfo {volume} {237}},\ \bibinfo
  {pages} {343} (\bibinfo {year} {1992})}\BibitemShut {NoStop}%
\bibitem [{\citenamefont {Machicoane}\ \emph {et~al.}(2018)\citenamefont
  {Machicoane}, \citenamefont {Labarre}, \citenamefont {Voisin}, \citenamefont
  {Moisy},\ and\ \citenamefont {Cortet}}]{Machicoane18}%
  \BibitemOpen
  \bibfield  {author} {\bibinfo {author} {\bibfnamefont {N.}~\bibnamefont
  {Machicoane}}, \bibinfo {author} {\bibfnamefont {V.}~\bibnamefont {Labarre}},
  \bibinfo {author} {\bibfnamefont {B.}~\bibnamefont {Voisin}}, \bibinfo
  {author} {\bibfnamefont {F.}~\bibnamefont {Moisy}},\ and\ \bibinfo {author}
  {\bibfnamefont {P.-P.}\ \bibnamefont {Cortet}},\ }\bibfield  {title}
  {\bibinfo {title} {Wake of inertial waves of a horizontal cylinder in
  horizontal translation},\ }\href@noop {} {\bibfield  {journal} {\bibinfo
  {journal} {Phys Rev Fluids}\ }\textbf {\bibinfo {volume} {3}} (\bibinfo
  {year} {2018})}\BibitemShut {NoStop}%
\bibitem [{\citenamefont {Johnson}(1982)}]{Johnson82a}%
  \BibitemOpen
  \bibfield  {author} {\bibinfo {author} {\bibfnamefont {E.}~\bibnamefont
  {Johnson}},\ }\bibfield  {title} {\bibinfo {title} {The effects of obstacle
  shape and viscosity in deep rotating flow over finite-height topography},\
  }\href@noop {} {\bibfield  {journal} {\bibinfo  {journal} {J Fluid Mech}\
  }\textbf {\bibinfo {volume} {120}},\ \bibinfo {pages} {359} (\bibinfo {year}
  {1982})}\BibitemShut {NoStop}%
\bibitem [{\citenamefont {Williamson}(1996)}]{Williamson1996}%
  \BibitemOpen
  \bibfield  {author} {\bibinfo {author} {\bibfnamefont {C.~H.}\ \bibnamefont
  {Williamson}},\ }\bibfield  {title} {\bibinfo {title} {Vortex dynamics in the
  cylinder wake},\ }\href@noop {} {\bibfield  {journal} {\bibinfo  {journal}
  {Ann. Rev. Fluid Mech.}\ }\textbf {\bibinfo {volume} {28}},\ \bibinfo {pages}
  {477} (\bibinfo {year} {1996})}\BibitemShut {NoStop}%
\bibitem [{\citenamefont {Douglas}\ and\ \citenamefont
  {Dupont}(1976)}]{DouglasDupont1976}%
  \BibitemOpen
  \bibfield  {author} {\bibinfo {author} {\bibfnamefont {J.}~\bibnamefont
  {Douglas}}\ and\ \bibinfo {author} {\bibfnamefont {T.}~\bibnamefont
  {Dupont}},\ }\bibfield  {title} {\bibinfo {title} {Interior penalty
  procedures for elliptic and parabolic {G}alerkin methods},\ }in\ \href@noop {}
  {\emph {\bibinfo {booktitle} {Comput Methods Appl Sci}}},\
  \bibinfo {editor} {edited by\ \bibinfo {editor} {\bibfnamefont
  {R.}~\bibnamefont {Glowinski}}\ and\ \bibinfo {editor} {\bibfnamefont
  {J.~L.}\ \bibnamefont {Lions}}}\ (\bibinfo  {publisher} {Springer},\  \bibinfo {year}
  {1976})\ pp.\ \bibinfo {pages} {207--216}\BibitemShut {NoStop}%
\bibitem [{\citenamefont {Burman}\ and\ \citenamefont
  {Hansbo}(2006)}]{BurmanHansbo2006}%
  \BibitemOpen
  \bibfield  {author} {\bibinfo {author} {\bibfnamefont {E.}~\bibnamefont
  {Burman}}\ and\ \bibinfo {author} {\bibfnamefont {P.}~\bibnamefont
  {Hansbo}},\ }\bibfield  {title} {\bibinfo {title} {Edge stabilization for the
  generalized stokes problem: a continuous interior penalty method},\
  }\href@noop {} {\bibfield  {journal} {\bibinfo  {journal} {Comput Methods
  Appl Mech Engrg}\ }\textbf {\bibinfo {volume} {195}},\
  \bibinfo {pages} {2393} (\bibinfo {year} {2006})}\BibitemShut {NoStop}%
\bibitem [{\citenamefont {Burman}\ \emph {et~al.}(2006)\citenamefont {Burman},
  \citenamefont {Fern{\'a}ndez},\ and\ \citenamefont
  {Hansbo}}]{Burmanetal2006}%
  \BibitemOpen
  \bibfield  {author} {\bibinfo {author} {\bibfnamefont {E.}~\bibnamefont
  {Burman}}, \bibinfo {author} {\bibfnamefont {M.}~\bibnamefont
  {Fern{\'a}ndez}},\ and\ \bibinfo {author} {\bibfnamefont {P.}~\bibnamefont
  {Hansbo}},\ }\bibfield  {title} {\bibinfo {title} {Continuous interior
  penalty finite element method for {O}seen's equations},\ }\href@noop {}
  {\bibfield  {journal} {\bibinfo  {journal} {SIAM J Numer Anal}\ }\textbf
  {\bibinfo {volume} {44}},\ \bibinfo {pages} {1248} (\bibinfo {year}
  {2006})}\BibitemShut {NoStop}%
\bibitem [{\citenamefont {Tong}\ \emph {et~al.}(2022)\citenamefont {Tong},
  \citenamefont {Kamensky},\ and\ \citenamefont {Evans}}]{Tongetal2022}%
  \BibitemOpen
  \bibfield  {author} {\bibinfo {author} {\bibfnamefont {G.~G.}\ \bibnamefont
  {Tong}}, \bibinfo {author} {\bibfnamefont {D.}~\bibnamefont {Kamensky}},\
  and\ \bibinfo {author} {\bibfnamefont {J.~A.}\ \bibnamefont {Evans}},\
  }\bibfield  {title} {\bibinfo {title} {Skeleton-stabilized
  divergence-conforming b-spline discretizations for incompressible flow
  problems of high {R}eynolds number},\ }\href@noop {} {\bibfield  {journal}
  {\bibinfo  {journal} {Comput \& Fluids}\ }\textbf {\bibinfo {volume}
  {248}},\ \bibinfo {pages} {105667} (\bibinfo {year} {2022})}\BibitemShut
  {NoStop}%
\bibitem [{\citenamefont {Moura}\ \emph {et~al.}(2022)\citenamefont {Moura},
  \citenamefont {Cassinelli}, \citenamefont {da~Silva}, \citenamefont
  {Burman},\ and\ \citenamefont {Sherwin}}]{Mouraetal2022}%
  \BibitemOpen
  \bibfield  {author} {\bibinfo {author} {\bibfnamefont {R.~C.}\ \bibnamefont
  {Moura}}, \bibinfo {author} {\bibfnamefont {A.}~\bibnamefont {Cassinelli}},
  \bibinfo {author} {\bibfnamefont {A.~F.}\ \bibnamefont {da~Silva}}, \bibinfo
  {author} {\bibfnamefont {E.}~\bibnamefont {Burman}},\ and\ \bibinfo {author}
  {\bibfnamefont {S.~J.}\ \bibnamefont {Sherwin}},\ }\bibfield  {title}
  {\bibinfo {title} {Gradient jump penalty stabilisation of spectral/hp element
  discretisation for under-resolved turbulence simulations},\ }\href@noop {}
  {\bibfield  {journal} {\bibinfo  {journal} {Comput Methods Appl
  Mech Engrg}\ }\textbf {\bibinfo {volume} {388}},\ \bibinfo
  {pages} {114200} (\bibinfo {year} {2022})}\BibitemShut {NoStop}%
\bibitem [{\citenamefont {Fern{\'a}ndez}(2013)}]{Fernandez2013}%
  \BibitemOpen
  \bibfield  {author} {\bibinfo {author} {\bibfnamefont {M.~A.}\ \bibnamefont
  {Fern{\'a}ndez}},\ }\bibfield  {title} {\bibinfo {title} {Incremental
  displacement-correction schemes for incompressible fluid-structure
  interaction: stability and convergence analysis},\ }\href@noop {} {\bibfield
  {journal} {\bibinfo  {journal} {Numer Math}\ }\textbf {\bibinfo
  {volume} {123}},\ \bibinfo {pages} {21} (\bibinfo {year} {2013})}\BibitemShut
  {NoStop}%
\bibitem [{\citenamefont {Burman}\ \emph {et~al.}(2020)\citenamefont {Burman},
  \citenamefont {Fern{\'a}ndez},\ and\ \citenamefont
  {Frei}}]{BurmanFernandezFrei2020}%
  \BibitemOpen
  \bibfield  {author} {\bibinfo {author} {\bibfnamefont {E.}~\bibnamefont
  {Burman}}, \bibinfo {author} {\bibfnamefont {M.~A.}\ \bibnamefont
  {Fern{\'a}ndez}},\ and\ \bibinfo {author} {\bibfnamefont {S.}~\bibnamefont
  {Frei}},\ }\bibfield  {title} {\bibinfo {title} {A {N}itsche-based
  formulation for fluid-structure interactions with contact},\ }\href@noop {}
  {\bibfield  {journal} {\bibinfo  {journal} {ESAIM: M2AN}\ }\textbf {\bibinfo {volume} {54}},\ \bibinfo {pages}
  {531} (\bibinfo {year} {2020})}\BibitemShut {NoStop}%
\bibitem [{\citenamefont {Jacobs}(1964)}]{Jacobs64a}%
  \BibitemOpen
  \bibfield  {author} {\bibinfo {author} {\bibfnamefont {S.}~\bibnamefont
  {Jacobs}},\ }\bibfield  {title} {\bibinfo {title} {On stratified flows over
  bottom topography},\ }\href@noop {} {\bibfield  {journal} {\bibinfo
  {journal} {J Mar Res}\ ,\ \bibinfo {pages} {223}} (\bibinfo {year}
  {1964})}\BibitemShut {NoStop}%
\bibitem [{\citenamefont {Hecht}(2013)}]{FreeFem}%
  \BibitemOpen
  \bibfield  {author} {\bibinfo {author} {\bibfnamefont {F.}~\bibnamefont
  {Hecht}},\ }\bibfield  {title} {\bibinfo {title} {New development in
  {F}ree{F}em++},\ }\href@noop {} {\bibfield  {journal} {\bibinfo  {journal} {J
  Numer Math}\ }\textbf {\bibinfo {volume} {20}},\ \bibinfo {pages} {251}
  (\bibinfo {year} {2013})}\BibitemShut {NoStop}%
\bibitem [{\citenamefont {Ayachit}(2015)}]{ParaView}%
  \BibitemOpen
  \bibfield  {author} {\bibinfo {author} {\bibfnamefont {U.}~\bibnamefont
  {Ayachit}},\ }\href@noop {} {\emph {\bibinfo {title} {The ParaView Guide: A
  Parallel Visualization Application}}}\ (\bibinfo  {publisher} {Kitware,
  Inc.},\ \bibinfo {address} {USA},\ \bibinfo {year} {2015})\BibitemShut
  {NoStop}%
\end{thebibliography}

\bibliographystyle{plain}
%

\end{document}